\documentclass[aip, cha, amsmath, amssymb, reprint]{revtex4-2}

\usepackage{xcolor}
\usepackage{stix2}

\usepackage{graphicx}
\usepackage{dcolumn}

\usepackage[utf8]{inputenc}
\usepackage[T1]{fontenc}
\usepackage[english]{babel}

\usepackage{mathtools}

\newcommand{\f}[2][\!]{#1\left(#2\right)}
\newcommand{\abs}[1]{\left|#1\right|}
\newcommand{\D}[1]{\,\mathrm{d}#1}
\newcommand{\Dr}[1]{\mathrm{d}#1\,}
\newcommand{\Dd}[1]{\mathrm{d}#1}
\newcommand{\I}{\mathrm{i}}
\newcommand{\E}[1]{\mathrm{e}^{#1}}
\newcommand{\sgn}{\mathop{\mathrm{sgn}}\nolimits}
\newcommand{\vp}{\mathop{\mathrm{v.p.}}\nolimits}

\renewcommand{\Im}{\mathop{\mathrm{Im}}\nolimits}
\newcommand{\const}{\mathop{\mathrm{const}}\nolimits}
\newcommand{\twolinebrackets}[4]{\left#1 \vphantom{#3#4}#3 \right. \\
  \left. \vphantom{#3#4}#4 \right#2}
\newcommand{\threelinebrackets}[5]{\left#1 \vphantom{#3#4#5}#3 \right. \\
  #4 \\
  \left. \vphantom{#3#4#5}#5 \right#2}

\newcommand{\fivelinebrackets}[7]{\left#1 \vphantom{#3#4#5#6#7}#3 \right. \\
  #4 \\
  #5 \\
  #6 \\
  \left. \vphantom{#3#4#5#6#7}#7 \right#2}

\newcommand{\Dg}{\mathop{\Delta g}\nolimits}
\newcommand{\De}{\mathop{\Delta\varepsilon}\nolimits}
\newcommand{\Dte}{\mathop{\Delta\teps}\nolimits}
\newcommand{\Ds}{\mathop{\Delta\sigma}\nolimits}
\newcommand{\Dts}{\mathop{\Delta\tilde\sigma}\nolimits}
\newcommand{\Da}{\mathop{\Delta\alpha}\nolimits}
\newcommand{\tomega}{{\tilde\omega}}
\newcommand{\tOmega}{{\tilde\Omega}}
\newcommand{\teps}{{\tilde\varepsilon}}
\newcommand{\tlt}{{\tilde t}}

\allowdisplaybreaks
\setcitestyle{numbers,square}
\begin{document}

%\preprint{}

\title[Synchronization transitions and sensitivity to asymmetry in the bimodal Kuramoto systems with Cauchy noise]{Synchronization transitions and sensitivity to asymmetry in the bimodal Kuramoto systems with Cauchy noise
}

\author{V. A. Kostin}
\affiliation{Gaponov-Grekhov Institute of Applied Physics of the Russian Academy of Sciences, 46 Ul'yanov Street, Nizhny Novgorod, 603950, Russia}
\affiliation{Department of Control Theory, Lobachevsky State University of Nizhny Novgorod, 23 Gagarin Avenue, Nizhny Novgorod, 603022, Russia}
\author{V. O. Munyaev}
\affiliation{Department of Control Theory, Lobachevsky State University of Nizhny Novgorod, 23 Gagarin Avenue, Nizhny Novgorod, 603022, Russia}
%\affiliation{Scientific and Educational Mathematical Center “Mathematics of Future Technologies”, Lobachevsky State University of Nizhny Novgorod, 23 Gagarin Avenue, Nizhny Novgorod, 603022, Russia}
\author{G. V. Osipov}
\affiliation{Department of Control Theory, Lobachevsky State University of Nizhny Novgorod, 23 Gagarin Avenue, Nizhny Novgorod, 603022, Russia}
%\affiliation{Scientific and Educational Mathematical Center “Mathematics of Future Technologies”, Lobachevsky State University of Nizhny Novgorod, 23 Gagarin Avenue, Nizhny Novgorod, 603022, Russia}
\author{L. A. Smirnov}
\affiliation{Department of Control Theory, Lobachevsky State University of Nizhny Novgorod, 23 Gagarin Avenue, Nizhny Novgorod, 603022, Russia}
%\affiliation{Scientific and Educational Mathematical Center “Mathematics of Future Technologies”, Lobachevsky State University of Nizhny Novgorod, 23 Gagarin Avenue, Nizhny Novgorod, 603022, Russia}

\date{\today}% It is always \today, today,

\begin{abstract}
  We analyze the synchronization dynamics of the thermodynamically large systems of globally coupled phase oscillators under Cauchy noise forcings with bimodal distribution of frequencies and asymmetry between two distribution components.
  The systems with the Cauchy noise admit the application of the Ott-Antonsen ansatz, which has allowed us to study analytically synchronization transitions both in the symmetric and asymmetric cases.
  The dynamics and the transitions between various synchronous and asynchronous regimes are shown to be very sensitive to the asymmetry degree whereas the scenario of the symmetry breaking is universal and does not depend on the particular way to introduce asymmetry, be it the unequal populations of modes in bimodal distribution, the phase delay of the Kuramoto–Sakaguchi model, the different values of the coupling constants, or the unequal noise levels in two modes.
  In particular, we found that even small asymmetry may stabilize the stationary partially synchronized state, and this may happen even for arbitrarily large frequency difference between two distribution modes (oscillator subgroups).
  This effect also results in the new type of bistability between two stationary partially synchronized states: one with large level of global synchronization and synchronization parity between two subgroups and another with lower synchronization where the one subgroup is dominant, having higher internal (subgroup) synchronization level and enforcing its oscillation frequency on the second subgroup.
  For the four asymmetry types, the critical values of asymmetry parameters were found analytically above which the bistability between incoherent and partially synchronized states is no longer possible.
\end{abstract}

\maketitle

\begin{quotation}
	Synchronization in ensembles of globally coupled phase oscillators with bimodal frequency distributions have been the subject of extensive study. Nowadays, researchers have already conducted a comprehensive analysis of this fundamental phenomenon in thermodynamically large systems.
  However, the influence of asymmetry between the distribution components and its impact on the system's behavior remain largely unexplored.
  In this work, we just focus on the presence of asymmetry between the distribution components. We thoroughly examine the effects of introducing various forms of asymmetry, such as non-equal subgroup populations, phase delays, and differing noise levels or coupling constants.
  The obtained results reveal that even subtle degrees of asymmetry can induce significant modifications in the system's dynamics, including the destruction of bistability and the emergence of asymmetric regimes where one oscillator subgroup dominates.
  Surprisingly, the observed scenario of symmetry breaking was found to be universal, unaffected by the specific type of asymmetry introduced.
  Additionally, the study uncovered a remarkable phenomenon when the introduction of asymmetry leads to stabilization of a stationary partially synchronized state, even in the presence of infinitely large frequency differences between the distribution modes.
  The corresponding stabilization effect also gives rise to a novel type of bistability where two partially synchronized states coexist, each exhibiting distinct levels of global synchronization and dominance of one subgroup.
  One of these two partially synchronized states has higher level of synchronization with dominance of one subgroup over another, while the other state demonstrates synchronization parity between the subgroups.
  Our study, which employed both macroscopic approaches based on the Ott–Antonsen ansatz and kinetic simulations, provides profound insights into the impact of asymmetry on dynamics of complex systems and ways to control of their properties.
  The present discoveries open new avenues for understanding the intricate interplay between asymmetry and synchronization and uncovers the universal nature of symmetry-breaking phenomena in collective behavior in complex oscillator systems with implications for diverse scientific and engineering applications.
\end{quotation}

\section{Introduction}

Synchronization in the oscillator ensembles is an important natural phenomenon occurring in various scientific and engineering areas~\cite{pikovsky_synchronization_2001, osipov_synchronization_2007, boccaletti_synchronization_2018}.
The oscillators with different natural frequencies or under action of noisy forcings may synchronize and oscillate at a common frequency.
The significant theoretical advance in understanding these processes is achieved using the Kuramoto model~\cite{kuramoto_self-entrainment_1975} and its extensions~\cite{acebron_kuramoto_2005, gupta_kuramoto_2014, pikovsky_dynamics_2015, kundu_transition_2017, gherardini_spontaneous_2018, kundu_perfect_2018, gong_repulsively_2019, kundu_synchronization_2019} that provide an abstract paradigmal description preserving major features of synchronization dynamics for the globally coupled oscillator systems of different physical or biological nature.
The distribution of oscillator over natural frequency is one of the key parameters strongly affecting the synchronization dynamics.
Though the majority of the studies assume unimodal frequency distributions, the bimodal distributions also attract considerable attention.

The dynamics of the oscillator systems with symmetric bimodal distribution was studied in details~\cite{bonilla_nonlinear_1992, crawford_amplitude_1994, perez_moment-based_1997, acebron_asymptotic_1998, bonilla_time-periodic_1998, acebron_synchronization_2000, martens_exact_2009, pazo_existence_2009, munyaev_analytical_2020, campa_phase_2020}.
In particular, in~\cite{martens_exact_2009}, the possible dynamical regimes and bifurcation between them were comprehensively analyzed using the Ott–Antonsen ansatz.
The corresponding dynamics is quite rich with region of bistability between asynchronous and synchronous states whereas synchronous state may be characterized either by a constant or oscillating order parameter.
What should be noted, the stable macroscopic dynamics is symmetric with respect to the two components of the bimodal distribution, and there are no stable regimes when one of two oscillator subgroups is dominant and imposes its natural frequency upon the other subgroup.
The introduction of asymmetry is known to enrich the dynamics generally, for example, to induce new types of multistability~\cite{kim_asymmetric_2011, pisarchik_asymmetry_2018, pisarchik_multistability_2022}.
The studies of asymmetric bimodal systems~\cite{bonilla_nonlinear_1992, acebron_asymptotic_1998, acebron_breaking_1998, montbrio_synchronization_2004, barreto_synchronization_2008, manoranjani_influence_2022} show that the dynamics and transitions may indeed be more complicated and richer.
However, the comprehensive analysis of the possible synchronization regimes in the thermodynamically large asymmetric systems has never been performed to our knowledge. 

The stationary partially synchronized states can be found analytically for an arbitrary frequency distribution using, for example, the idea of subgroup order parameter~\cite{omelchenko_nonuniversal_2012, munyaev_analytical_2020}.
However, to study the non-stationary regimes analytically as well, one may require additional assumptions about the frequency distributions.
In particular, for systems with the Lorentzian distribution of natural frequencies, the Ott–Antonsen ansatz is widely used~\cite{ott_low_2008}.
It allows one to reduce the dimensionality of the one-particle kinetic Fokker–Planck-type equation and study the dynamics of the order parameter directly.
In~\cite{martens_exact_2009}, it was used to study the noiseless case with two components of the bimodal distribution having the Lorentzian shapes by reducing the kinetic equation to the ordinary differential equations for two subgroup order parameters corresponding to the two Lorentzian modes of the frequency distribution.
Recently, in~\cite{tonjes_low-dimensional_2020}, it was shown that the systems with the Cauchy noise also allow the direct application of the Ott–Antonsen ansatz.
Such Cauchy noise may arise, for example, from the Poisson process of the delta pulses with the wrapped Cauchy distribution of amplitudes.

Here, we employ this idea for the Cauchy noise to study asymmetric systems with bimodal frequency distributions and analyze how introducing asymmetry in the large (i.e., in the thermodynamic limit) systems with bimodal distribution breaks the symmetry of the macroscopic dynamics for different types of asymmetry: unequal populations of two oscillator subgroups, the non-zero phase delay in the Kuramoto–Sakaguchi model, the unequal noise levels or coupling constants for the two subgroups.
Using both the macroscopic approaches based on the Ott–Antonsen ansatz and kinetic simulations of Fokker–Planck-type equations, we found that even slight asymmetry may significantly modify the dynamics of the system and the corresponding map of the dynamical regimes, for example, by destroying bistability or introducing new significantly asymmetric dynamical regimes where one of the subgroups is dominant (symmetry breaking).
The scenario of how the map of dynamical regimes changes with introducing the asymmetry turns out to be universal and does not depend on the particular type of the asymmetry.

The article is organized as follows.
In Sec.~\ref{sec:2}, we introduce the considered model and perform the reduction of dimensionality using the Ott–Antonsen ansatz.
Section~\ref{sec:3} summarizes the previous results for the symmetric case as well as describes the unstable asymmetric states in the symmetric systems that will play their roles in the asymmetric case.
A comprehensive description of the synchronization transitions, respective bifurcations, and dynamical regimes for the case with asymmetry of subgroup populations is given in Sec.~\ref{sec:4}.
Section~\ref{sec:5} briefly explains how the obtained results for the population asymmetry change for other kinds of asymmetry.
We conclude with Sec.~\ref{sec:6}.
Appendix contains the analytical asymptotic formulas for the parameters of partially syncronized states and the positions of their bifurcations.

\section{Model and equations}
\label{sec:2}

We consider the Kuramoto–Sakaguchi system 
\begin{gather}
  \label{eq:initial}
  \dot\varphi_j = \omega_j + \frac {\varepsilon_j}N \sum_{j' = 1}^N\sin\f{\varphi_{j'} - \varphi_j - \alpha_j} + \zeta_j
\end{gather}
for $N$ coupled noisy phase oscillators with phases $\varphi_j(t)$, frequencies $\omega_j$, coupling constants $\varepsilon_{j}$, constant phase delays $\alpha_{j}$, and the mutually independent Cauchy noise forces $\zeta_j(t)$.
Here, we analyze the thermodynamic limit $N \to \infty$ assuming some given distribution of frequencies $\omega_j$, coupling coefficients $\varepsilon_j \geqslant 0$, phase delays $\alpha_j$, and noise powers $\sigma_j \geqslant 0$ with joint density $g(\omega, \sigma, \varepsilon, \alpha)$.
This joint density gives the probability $\D{P} = g(\omega, \varepsilon, \alpha, \sigma) \D{\omega} \D{\varepsilon} \D{\alpha} \D{\sigma}$ of a randomly chosen oscillator $j$ to have the frequency $\omega_j \in (\omega, \omega + \D{\omega})$, the coupling constant $\varepsilon_j \in (\varepsilon, \varepsilon + \D{\varepsilon})$, the delay $\alpha_j \in (\alpha, \alpha + \D{\alpha})$, and the noise power $\sigma_j \in (\sigma, \sigma + \D{\sigma})$.
This joint density satisfies the normalization condition
\begin{gather*}
  \label{eq:norm}
  \int_{-\infty}^{\infty}\Dr{\omega} \int_{0}^{\infty}\Dr{\varepsilon} \int_{-\pi}^{\pi}\Dr{\alpha} \int_{0}^{\infty}\Dr{\sigma} g\f{\omega, \varepsilon, \alpha, \sigma} = 1.
\end{gather*}
In what follows we consider the discrete bimodal distribution with two subgroups of oscillators, the subgroup $m$ having frequency $\hat\omega_m$, coupling constant $\hat\varepsilon_m$, phase delay $\hat\alpha_m$, and noise power $\hat\sigma_m$, $m = 1, 2$, so that
\begin{multline*}
  g\f{\omega, \varepsilon, \alpha, \sigma} \\
  {}= \sum_{m = 1}^2 g_m \delta\f{\omega - \hat\omega_m} \delta\f{\varepsilon - \hat\varepsilon_m} \delta\f{\alpha - \hat\alpha_m} \delta\f{\sigma - \hat\sigma_m},
\end{multline*}
where $0 \leqslant g_m \leqslant 1$ are the amount fractions of oscillators in subgroups with the normalization condition $g_1 + g_2 = 1$.
We could have considered the Lorentzian frequency distributions instead of the delta functions in the two subgroups following~\cite{martens_exact_2009}.
This would not have changed the form of the resulting reduced equations for the order parameters of the two subgroups, but only effectively modified the noise powers $\hat\sigma_{1,2}$ depending on the widths of the Lorentzians.
Therefore, to keep the derivation simpler as well as to be able to provide more straightforward comparison with the case of the Gaussian noise in~\cite{,acebron_asymptotic_1998,  acebron_breaking_1998}, we limit ourselves to the discrete case.

The initial equation~\eqref{eq:initial} can also be rewritten as
\begin{gather*}
  \dot\varphi_j = \omega_j + \varepsilon_j \Im\f{r \E{-\I\varphi_j - \I\alpha_j}} + \zeta_j,
\end{gather*}
where
\begin{gather*}
  r = \frac 1N \sum_{j = 1}^N \E{\I\varphi_{j}}
\end{gather*}
is the complex mean field or the complex order parameter.
With the multi-oscillator correlations assumed insignificant, the time-dependent single-oscillator distribution functions $p_{1, 2}(\varphi, t)$ over phases of the subgroups satisfies the Fokker–Planck-type equations
\begin{gather}
  \label{eq:fokker_planck}
  \frac{\partial p_m}{\partial t} + \frac{\partial}{\partial \varphi}\left\{p_m\left[\hat\omega_m + \hat\varepsilon_m\Im\f{R \E{-\I\varphi - \I\hat\alpha_m}}\right] \right\} = \hat\sigma_m \hat Cp_m,
\end{gather}
where
\begin{gather*}
  R = g_1 Z_1 + g_2 Z_2
\end{gather*}
is the global order parameter in the thermodynamic limit ($r \to R$ at $N \to \infty$), $Z_m \equiv z_m^{(1)}$ is the subgroup order parameter,
\begin{gather*}
  z_m^{(k)} = \frac 1{2\pi} \int_{-\pi}^{\pi} p_m \E{\I k \varphi} \D{\varphi}
\end{gather*}
are the subgroup circular moments, and $\hat C$ is the Cauchy noise operator,
\begin{multline*}
  \hat Cp_m = -\sum_{k = -\infty}^{\infty} \abs{k}z_m^{(k)} \E{-\I k\varphi} \\
  {}= \frac 1{2\pi} \frac{\partial}{\partial\varphi} \vp \int_{-\pi}^{\pi} p_m\f{\varphi', t} \cot\frac{\varphi' - \varphi}2 \D{\varphi'}.
\end{multline*}

The Fokker–Planck equation~\eqref{eq:fokker_planck} can be rewritten in representation of circular moments $z_{1, 2}^{(k)}$ as the following infinite set of the coupled ordinary differential equations:
\begin{multline}
  \label{eq:kinetic}
  \dot z_m^{(k)} = k\left[\I \hat\omega_m z_m^{(k)} + \frac{\hat\varepsilon_m}2 \left(\E{-\I\hat\alpha_m} R z_{m}^{(k - 1)} - \E{\I\hat\alpha_m} R^{*} z_{m}^{(k + 1)}\right)\right] \\
  {}- \abs{k}\hat\sigma_m z_m^{(k)},
\end{multline}
where the dot on top denotes the time derivative, and the star denotes the complex conjugation.
This equation set admits the Ott–Antonsen ansatz $z_m^{(k)} = Z_m^k$ and $z_{m}^{(-k)} = (Z_m^{*})^k$ for $k \geqslant 0$ which results in the finite equation set for the complex subgroup order parameter $Z_{1, 2}$,
\begin{gather}
  \label{eq:ott-attonsen}
  \dot Z_m = \left(\I \hat\omega_m - \hat\sigma_m\right) Z_m + \frac{\hat\varepsilon_m}2 \left(\E{-\I\hat\alpha_m} R - \E{\I\hat\alpha_m} R^{*} Z_m^2\right).
\end{gather}

In what follows, we use the following notation that separates out the asymmetry of the two oscillator subgroups: $\hat\omega_{1, 2} = \Omega \mp \omega_0$, $g_{1, 2} = (1 \mp \Dg)/2$, $\hat\sigma_{1, 2} = \sigma_0 (1 \mp \Ds)$, $\hat\varepsilon_{1, 2} = \varepsilon_0 (1 \mp \De)$, $\hat\alpha_{1, 2} = \Da \mp \alpha_0$, where $\Omega = (\hat\omega_1 + \hat\omega_2)/2$, $\sigma_0 = (\hat\sigma_1 + \hat\sigma_2)/2$, and $\varepsilon_0 = (\hat\varepsilon_1 + \hat\varepsilon_2)/2$ are the median values of the natural frequency, noise power, and coupling coefficient, respectively, $\alpha_0 = (\hat\alpha_2 - \hat\alpha_1)/2$ is the symmetric phase delay, $\omega_0 = (\hat\omega_2 - \hat\omega_1)/2$ is the halved frequency difference of the two subgroups, and $\Dg = g_2 - g_1$, $\Ds = (\hat\sigma_2 - \hat\sigma_1)/(\hat\sigma_2 + \hat\sigma_1)$, $\De = (\hat\varepsilon_2 - \hat\varepsilon_1)/(\hat\varepsilon_2 + \hat\varepsilon_1)$, and $\Da = (\hat\alpha_1 + \hat\alpha_2)/2$ are the asymmetry measures of the subgroup populations, noise powers, coupling coefficients, and phase delays, respectively.
For the definiteness, we assume $\omega_0 \geqslant 0$.
Everywhere below, we also assume $\abs{\Da} < \pi/2$, $\sigma_0 > 0$.

By substitution $Z_{1, 2} = \E{-\I\hat\alpha_{1,2} + \I\Omega t}\mathcal Z_{1, 2}$, $R = \mathcal R \E{\I\Omega t}$, one arrives from Eqs.~\eqref{eq:ott-attonsen} at the equation set
\begin{gather}
  \label{eq:Z1}
  \dot{\mathcal Z_1} = \left(-\I\omega_0 + \Ds\sigma_0 - \sigma_0\right) \mathcal Z_1 + \frac{\varepsilon_0}2 \left(1 - \De\right) \left(\mathcal R - \mathcal R^{*} \mathcal Z_1^2\right), \\
  \label{eq:Z2}
  \dot{\mathcal Z_2} = \left(\I\omega_0 - \Ds\sigma_0 - \sigma_0\right) \mathcal Z_2 + \frac{\varepsilon_0}2 \left(1 + \De\right) \left(\mathcal R - \mathcal R^{*} \mathcal Z_2^2\right), \\
  \label{eq:R}
  \mathcal R = \frac{\E{-\I\Da}}2 \left[\left(1 - \Dg\right)\E{\I\alpha_0}\mathcal Z_1 + \left(1 + \Dg\right)\E{-\I\alpha_0}\mathcal Z_2\right].
\end{gather}
Below, we investigate possible stable dynamical regimes of Eqs.~\eqref{eq:Z1}–\eqref{eq:R} for $\alpha_0 = 0$ analytically and numerically.
After time normalization $t = \tlt/\sigma_0$, $\omega_0 = \sigma_0 \tomega$, $\varepsilon_0 = \sigma_0 \teps$, Eqs.~\eqref{eq:Z1}–\eqref{eq:Z2} are left with 6 essential parameters:
\begin{gather}
  \label{eq:normalized-Z1}
  \frac{\Dd{\mathcal Z_1}}{\Dd{\tlt}} = \left(-\I\tomega + \Ds - 1\right) \mathcal Z_1 + \frac{\teps}2 \left(1 - \De\right) \left(\mathcal R - \mathcal R^{*} \mathcal Z_1^2\right), \\
  \label{eq:normalized-Z2}
  \frac{\Dd{\mathcal Z_2}}{\Dd{\tlt}} = \left(\I\tomega - \Ds - 1\right) \mathcal Z_2 + \frac{\teps}2 \left(1 + \De\right) \left(\mathcal R - \mathcal R^{*} \mathcal Z_2^2\right), \\
  \label{eq:normalized-R}
  \mathcal R = \E{-\I\Da} \frac{\left(1 - \Dg\right)\mathcal Z_1 + \left(1 + \Dg\right)\mathcal Z_2}2. 
{}\end{gather}
In what follows, we consider and compare the cases where no more than one of asymmetry parameters ($\Dg$, $\Ds$, $\De$, and $\Da$) is nonzero.

The initial equations~\eqref{eq:initial} as well as Eqs.~\eqref{eq:normalized-Z1}–\eqref{eq:normalized-R} are invariant with respect to the global phase shift $\varphi_j \mapsto \varphi_j + \beta$, $\mathcal Z_{1, 2} \mapsto \mathcal Z_{1, 2} \E{\I\beta}$, $\mathcal R \mapsto \mathcal R \E{\I\beta}$.
Therefore, all regimes in Eqs.~\eqref{eq:normalized-Z1}–\eqref{eq:normalized-R} (except for the fully incoherent state $\mathcal Z_{1, 2} = 0$) form families whose the members differ only by the constant global phase shift.
In what follows, we consider such families as a single regime without additional remarks.

The invariance with respect to the global phase also allows one to reduce the dimensionality of the system~\eqref{eq:normalized-Z1}–\eqref{eq:normalized-R} from 4 to 3.
To this end, one may employ polar coordinates $\mathcal Z_{1, 2}(t) = \rho_{1, 2}(t)\E{\I\theta_{1, 2}(t)}$ and consider $\rho_1$, $\rho_2$, and $\psi = \theta_2 - \theta_1$ as new real-valued dynamical variables.
These variables satisfy the equations
\begin{gather}
  \label{eq:rho1}
  \begin{multlined}
    \frac{\Dd{\rho_1}}{\Dd{\tlt}} = -\left(1 - \Ds\right) \rho_1 + \frac{\teps}4 \left(1 - \De\right) \left(1 - \rho_1^2\right) \\
    {}\times \left[\left(1 - \Dg\right) \rho_1 \cos\Da + \left(1 + \Dg\right) \rho_2 \cos\f{\psi - \Da}\right],
  \end{multlined} \\
  \label{eq:rho2}
  \begin{multlined}
    \frac{\Dd{\rho_2}}{\Dd{\tlt}} = -\left(1 + \Ds\right) \rho_2 + \frac{\teps}4 \left(1 + \De\right) \left(1 - \rho_2^2\right) \\
    {}\times \left[\left(1 - \Dg\right) \rho_1 \cos\f{\psi + \Da} + \left(1 + \Dg\right) \rho_2 \cos\Da\right],
  \end{multlined} \\
  \label{eq:psi}
  \begin{multlined}
    \frac{\Dd{\psi}}{\Dd{\tlt}} = 2\tomega - \frac{\teps}4 \left(1 - \De\right) \left(1 + \rho_1^2\right) \\
    {}\times \left[-\left(1 - \Dg\right) \sin\Da + \left(1 + \Dg\right) \frac{\rho_2}{\rho_1} \sin\f{\psi - \Da}\right] \\
    {} -\frac{\teps}4 \left(1 + \De\right) \left(1 + \rho_2^2\right) \\
    {}\times \left[\left(1 - \Dg\right) \frac{\rho_1}{\rho_2} \sin\f{\psi + \Da} + \left(1 + \Dg\right) \sin\Da\right].
  \end{multlined}
\end{gather}
Equations~\eqref{eq:rho1}–\eqref{eq:psi} may be convenient for studying the partially synchronized states and periodic regimes with $\rho_{1, 2} \neq 0$, whereas Eqs.~\eqref{eq:normalized-Z1}–\eqref{eq:normalized-R} are more suitable to study the asynchronous states and the associated bifurcations.
In particular, the stationary partially synchronized states are represented by the fixed points of system~\eqref{eq:rho1}–\eqref{eq:psi}.
These fixed points correspond to $\mathcal Z_{1, 2} = \rho_{1, 2} \E{\I\Omega_s t + \I\beta}$ with constant $\rho_{1, 2}$, $\Omega_s$, and $\beta$, where $\Omega_s = \sigma \tOmega_s$ is the difference between the synchronization (mean-field) frequency and the median frequency $\Omega$.

\section{Symmetric case}
\label{sec:3}

\begin{figure*}{}
  \includegraphics[width=\textwidth]{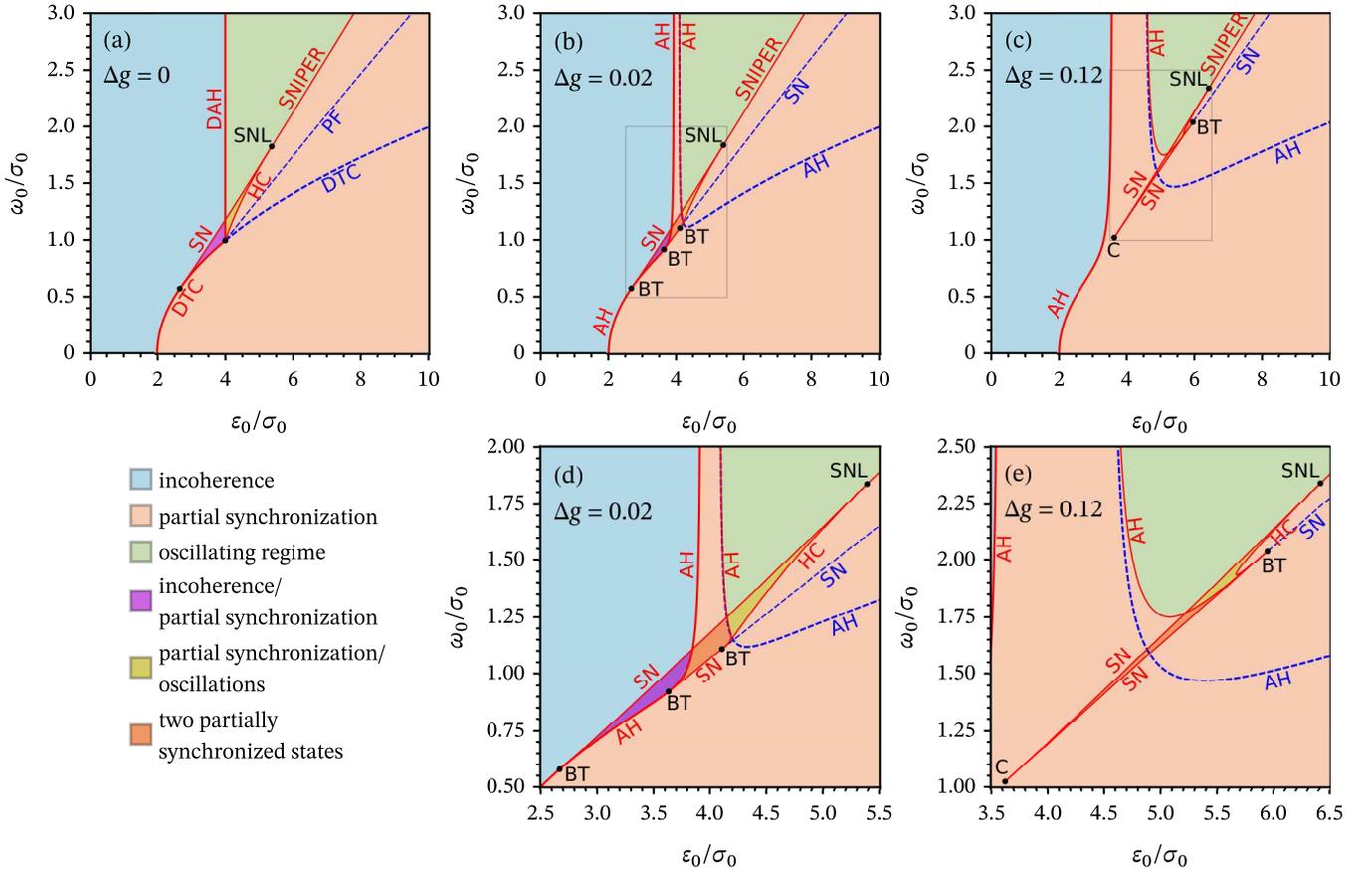}
  \caption{\label{fig:1} Bifurcation maps for the low-dimensional system~\eqref{eq:Z1}–\eqref{eq:R} describing dynamics of subgroup order parameters in the large ensemble of phase oscillators with the Cauchy noise and the discrete bimodal distribution of natural frequencies.
  The horizontal and vertical coordinates are, respectively, the coupling constant $\varepsilon_0$ and the halved subgroup frequency difference $\omega_0$ normalized by the Cauchy noise level $\sigma_0$.
  The different panels correspond to the different population asymmetries $\Dg$: (a) $0$, (b, d) $0.02$, (c, e) $0.12$.
  Other kinds of asymmetry are absent, $\De = \Ds = \Da = 0$.
  Panels (d) and (e) present the enlarged fragments (bounded by gray rectangles) of maps in (b) and (c), respectively.
  The red solid lines mark the bifurcations where the stable regimes are born or disappear; the blue dashed lines denote the bifurcations of stationary states where only unstable regimes participate.
  The dots indicate the bifurcations of higher codimension.
  The thicker lines and dots correspond to the degenerate bifurcations of the incoherent state.
  The colored areal shading denotes the possible dynamical regimes, as indicated in the legend.
  The following abbreviations are used for the bifurcation types: DAH, degenerate Andronov–Hopf; DTC, degenerate transcritical; AH, Andronov–Hopf; SN, saddle-node; SNIPER, saddle-node infinite-period; HC, homoclinic; PF, pitchfork; SNL, saddle-node-loop; BT, Bogdanov–Takens; C, cusp.}
\end{figure*}

\begin{figure*}{}
  \includegraphics[width=\textwidth]{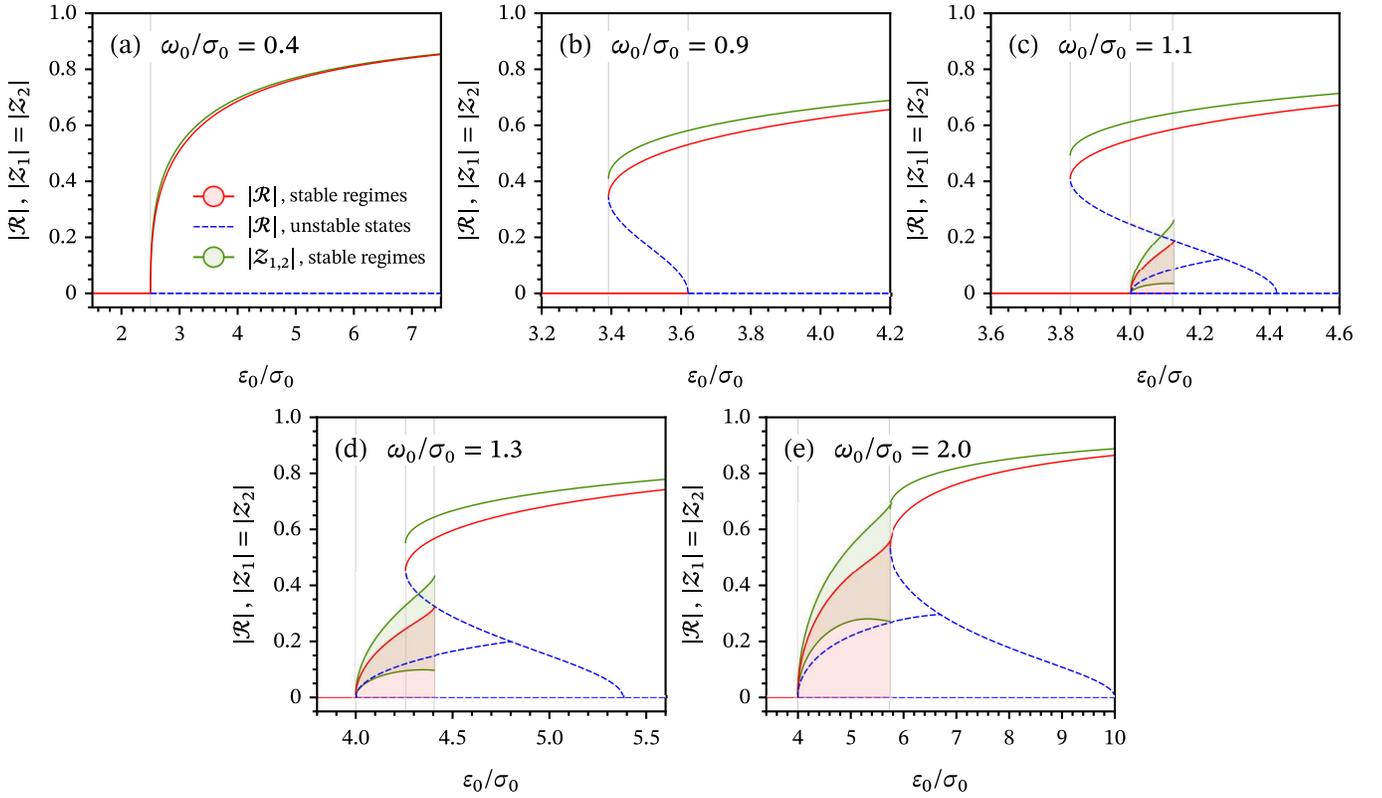}
  \caption{\label{fig:2} Synchronization branches showing the dependences of the global $\abs{\mathcal{R}}$ and subgroup $\abs{\mathcal{Z}_{1, 2}}$ order parameters on the normalized coupling strength $\varepsilon_0/\sigma_0$ in the symmetric case $\Dg = \De = \Ds = \Da = 0$ for different normalized subgroup frequency differences $\omega_0/\sigma_0$: (a) $0.4$, (b) $0.9$, (c) $1.1$, (d) $1.3$, and (e) $2.0$.
  For the stable stationary states, the red and green solid lines correspond to the global $\abs{\mathcal{R}}$ and subgroup $\abs{\mathcal{Z}_{1, 2}}$ order parameters, respectively.
  For the periodic regimes, the solid lines denote the minimal and maximal (over period) values of the order parameter and the area between them is shaded.
  The blue dashed lines indicate the value of the global order parameter $\abs{\mathcal{R}}$ in the unstable stationary states.
  The gray vertical lines mark the bifurcations of the stable regimes.}
\end{figure*}

The symmetric situation for Eqs.~\eqref{eq:normalized-Z1}–\eqref{eq:normalized-R} with $\Dg = \De =  \Ds = \Da = 0$ was studied in detail in~\cite{martens_exact_2009} (there, the parameter in place of $\sigma_0$ was the width of the Lorentzian, while here it is the power of the Cauchy noise).
In this section, we give a brief review of the known results with some additional remarks.
These results are illustrated by Fig.~\ref{fig:1}(a) where the map of the possible stable dynamical regimes is shown in parameter plane $(\teps, \tomega)$ and Fig.~\ref{fig:2} where the synchronization branches $\abs{\mathcal R(\teps)}, \abs{\mathcal Z_{1,2}(\teps)}$ are plotted for various values of $\tomega$.
All the stable regimes here are symmetric with $\abs{\mathcal Z_1} = \abs{\mathcal Z_2}$, $\mathcal Z_{1, 2}(\tlt) = \rho(\tlt) \E{\I\psi_0 \mp \I \psi(\tlt)/2}$, where $\rho$ and $\psi$ are real-valued time functions, and $\psi_0$ is a real constant which may be set to zero without loss of generality.
The functions $\rho$ and $\psi$ satisfy the equations
\begin{gather}
  \label{eq:symmetric-rho}
  \frac{\Dd{\rho}}{\Dd{\tlt}} = -\rho + \frac{\teps}4 \rho \left(1 - \rho^2\right) \left(1 + \cos\psi\right), \\
  \label{eq:symmetric-psi}
  \frac{\Dd{\psi}}{\Dd{\tlt}} = 2\tomega - \frac{\teps}2 \left(1 + \rho^2\right) \sin\psi.
\end{gather}
This system may support 3 types of stable regimes: asynchronous (incoherent) state, partially synchronized state, and oscillatory regime with partial synchronization (so-called standing wave).

\paragraph{Incoherent state.}
The stability analysis of the asynchronous (incoherent) state $\rho = \mathcal Z_{1, 2} = \mathcal R = 0$ can be performed through direct linearization of Eqs.~\eqref{eq:normalized-Z1}–\eqref{eq:normalized-R}.
The asynchronous state is stable at $\teps < \min\left\{4, 2 + 2\tomega^2\right\}$.
At $\tomega < 1$, $\teps = 2 + 2\tomega^2$, the asynchronous state loses its stability through the degenerate transcritical bifurcation, and, at $\tomega > 1$, $\teps = 4$, the degenerate Andronov–Hopf bifurcation occurs.

Note that the properties of these codimension-2 bifurcation may differ significantly from the common transcritical and Andronov–Hopf bifurcations.
So, in the degenerate supercritical Andronov–Hopf bifurcation, not 1, but 3 limit cycles (1 stable cycle and 2 unstable ones corresponding to unstable stationary partially synchronized states) are born along with incoherent state losing stability.
Therefore, this degenerate bifurcation may resemble the pitchfork and Andronov–Hopf bifurcations in some manner.
The degenerate transcritical bifurcation not only contains the change of the stability of the incoherently but actually involves creation of a circle of fixed points, and is also more similar to the pitchfork bifurcation.

\paragraph{Partially synchronized states.}
The stable symmetric partially synchronized state with $\rho = \const \neq 0$, $\psi = \const$ can be found by analyzing Eqs.~\eqref{eq:symmetric-rho} and \eqref{eq:symmetric-psi}.
By equaling the right-hand sides to zero, eliminating $\rho$, and using the Weierstrass substitution, one obtains the cubic equation
\begin{gather}
  \label{eq:cubic-psi}
  \left(\tau^2 + 1\right) \left(\tau + \tomega\right) - \teps \tau = 0
\end{gather}
for $\tau = \tan(\psi/2)$ in the partially synchronized equilibrium states of Eqs.~\eqref{eq:normalized-Z1}–\eqref{eq:normalized-R}; the equilibrium value of the order parameter is expressed as
\begin{gather}
  \abs{\mathcal R} = \rho \abs{\cos\frac{\psi}2} = \frac{\rho}{\sqrt{1 + \tau^2}}, \\
  \rho = \sqrt{\frac{\tomega - \tau}{\tomega + \tau}} = \sqrt{1 - \frac{2\left(1 + \tau^2\right)}{\teps}}
\end{gather}
and satisfies the bicubic equation
\begin{gather}
  \label{eq:cubic-varrho-1}
  \left(\varrho + 1\right)^2 \left(\varrho + 2 - \teps\right) + \tomega^2 \left(\varrho + 2\right) = 0,
\end{gather}
where $\varrho = \teps \abs{\mathcal R}^2$.

Equation~\eqref{eq:cubic-psi} has one real root if $D < 0$ and three real roots if $D > 0$, where
\begin{gather*}
  D = -4\tomega^4 + \left(\teps^2 - 20 \teps - 8\right) \tomega^2 + 4\left(\teps - 1\right)^3
\end{gather*}
is the discriminant of the respective cubic polynomial.
Among those roots, at $\tomega \leqslant 1/\sqrt{3}$, $\teps < 2 + 2\tomega^2$ or at $\tomega \geqslant 1/\sqrt{3}$, $D < 0$, no root gives $\rho$ between 0 and 1, so no partially synchronized states are possible.
At $\tomega \leqslant 1/\sqrt{3}$, $\teps > 2 + 2\tomega^2$, there exists one stable partially synchronized state which is born through abovementioned degenerate pitchfork bifurcation at $\teps = 2 + 2 \tomega^2$.
At $\tomega > 1/\sqrt{3}$, $D > 0$, there exists one stable partially synchronized state, and, if $\teps < 2 + 2\tomega^2$, there is additionally one unstable partially synchronized state.
A pair of stable and unstable states are born through saddle-node or saddle-node infinite-period bifurcation at $D = 0$, and the unstable state disappears at $\teps = 2 + 2\tomega^2$ through pitchfork bifurcation.
In the parameter plane $(\teps, \tomega)$, the segment of line $D = 0$ between points $(\teps, \tomega) = (8/3, 1/\sqrt{3})$ and $(\teps, \tomega) = (\teps_{\mathrm{SNL}}, \tomega_{\mathrm{SNL}}) \approx (5.3647, 1.8225)$ corresponds to the saddle-node bifurcation.
Here, $\teps_{\mathrm{SNL}}$ and $\tomega_{\mathrm{SNL}}$ denote the position of the saddle-node-loop bifurcation of codimension 2, where the saddle-node homoclinic trajectory is formed; the position of saddle-node in this bifurcation is $\rho = \rho_{\mathrm{SNL}} \approx 0.6493$, $\psi = \psi_{\mathrm{SNL}} \approx 1.2774$.
The semi-infinite segment starting at point $(\teps, \tomega) = (\teps_{\mathrm{SNL}}, \tomega_{\mathrm{SNL}})$ corresponds to the saddle-node infinite-period bifurcation.
Equation $D = 0$ is biquadratic in $\tomega$ and cubic in $\teps$, so it can be resolved to obtain analytical expressions for bifurcation values
\begin{gather}
  \notag
  \tomega = \tomega_{\mathrm{SN}}\f{\teps} \equiv \frac{\sqrt{2}}4 \sqrt{\teps^2 - 20\teps - 8 + \sqrt{\teps \left(\teps + 8\right)^3}}, \\
  \label{eq:eps-sn}
  \begin{multlined}
    \teps = \teps_{\mathrm{SN}}\f{\tomega} \equiv 1 - \frac{\tomega^2}{12} + \frac{\tomega^{2/3}}{12} \\
    {}\times \twolinebrackets{(}{)}{\sqrt[3]{5832 + 540\tomega^2 - \tomega^4 + 24\sqrt{3\left(27 - \tomega^2\right)^3}}}{{}+ \sqrt[3]{5832 + 540\tomega^2 - \tomega^4 - 24\sqrt{3\left(27 - \tomega^2\right)^3}}}
  \end{multlined}
\end{gather}
and order parameters $\rho = \sqrt{2 - \sqrt{1 + 8/\teps}}$, $\abs{\mathcal R} = (1/2) \sqrt{1 - 8/\teps + \sqrt{1 + 8/\teps}}$.
The parametric expressions for the bifurcation values in terms of the phase difference $\psi$ are more concise:
\begin{gather*}
  \tomega = \frac{2\tau^3}{1  - \tau^2} = \tan^2 \frac{\psi}2 \tan \psi, \\
  \teps = \frac{\left(1 + \tau^2\right)^2}{1 - \tau^2} = \frac{1}{\cos^2 \frac{\psi}2 \cos\psi}, \\
  \rho = \sqrt{\frac{3\tau^2 - 1}{1 + \tau^2}} = \sqrt{1 - 2 \cos\psi},
\end{gather*}
with $\pi/3 < \psi < \pi/2$, $1/\sqrt{3} < \tau < 1$.
Appendix gives asymptotic expressions for the position of the described bifurcations as well as for the parameters of the partially synchronized states.
In particular, the asymptotic behavior of the abovedescribed saddle-node bifurcation is given in Appendix~\ref{app:symmetric:sn}. 

Note that there may also be a pair of asymmetric unstable partially synchronized states with $\rho_1 \neq \rho_2$ in this system.
These asymmetric states are always unstable in the symmetric oscillator systems.
However, as seen from analysis in the next section, even slight asymmetry may lead to the stabilization of one of these partially synchronized states.
Therefore, we give a brief characterization of these states here.
These states are born in the same degenerate Andronov–Hopf bifurcation at $\tomega > 1$, $\teps = 4$ where the incoherent state loses stability and the oscillatory regime (standing wave described below) is born.
Thus, in the degenerate Andronov–Hopf bifurcation at $\tomega > 1$, $\teps = 4$ not one, but three limit cycles are formed.
However, these two unstable partially synchronized states are represented by limit cycles for the system~\eqref{eq:normalized-Z1}–\eqref{eq:normalized-Z2} only, but not for the reduced system~\eqref{eq:rho1}–\eqref{eq:psi}.
For the latter, these states are just two unstable fixed points.
Similarly to the symmetric state, the order parameter satisfies the bicubic equation
\begin{gather}
  \label{eq:cubic-varrho-2}
  \left(\varrho + 1\right) \left(3\varrho + 4 - \teps\right)^2 + \tomega^2 \left(4\varrho + 4 - \teps\right) = 0
\end{gather}
with $\varrho = \teps \abs{\mathcal R}^2$.
This equation can be obtained in different ways, for example, from the more general equation of the ninth degree in $\varrho$ which is described in the Sec.~\ref{sec:4} for asymmetric systems.
In the symmetric case, that equation splits into 3 cubic equations for $\varrho$: Eq.~\eqref{eq:cubic-varrho-1} and two equations of form~\eqref{eq:cubic-varrho-2}.
At $\tomega > 1$, $4 < \teps < \teps_{\mathrm{PF}}(\tomega)$, Eq.~\eqref{eq:cubic-varrho-2} has one solution with $0 < \varrho < \teps$ which represents two asymmetric unstable states where one of the oscillator subgroups dominates and forces its frequency onto the other one.
Here $\teps_{\mathrm{PF}}(\tomega)$ denotes the bifurcation value of $\teps$ where these asymmetric states disappear.
This bifurcation value lies between the respective values for the saddle-node and the degenerate transcritical bifurcation, $\teps_{\mathrm{SN}}(\tomega) < \teps_{\mathrm{PF}}(\tomega) < 2 + 2\tomega^2$. 
At the degenerate Andronov–Hopf bifurcation, $\tomega > 1$, $\teps \to 4 + 0$, these states have $\abs{\mathcal R}, \rho_{1, 2} \to 0$, and rotation frequency $\tOmega_s \equiv \Omega_s/\sigma_0 \to \pm \sqrt{\tomega^2 - 1}$, $\rho_2/\rho_1 \to \tomega \pm \sqrt{\tomega^2 - 1}$, $\psi \to \pi/2$.
With further increase in $\teps$, these states become more symmetric while $\tOmega_s$ and $\abs{\ln(\rho_2/\rho_1)}$ decrease.
Appendix~\ref{app:symmetric:couplings} describes this asymptotic behavior in more detail.

At $\teps = \teps_{\mathrm{PF}}(\omega)$, a pitchfork bifurcation happens where these two states become symmetric and merge with the unstable symmetric state, disappearing.
Note that the unstable symmetric state does not change the stability status at $\teps = \teps_{\mathrm{PF}}(\tomega)$.
Note also that in the initial system~\eqref{eq:normalized-Z1}–\eqref{eq:normalized-R}, this pitchfork bifurcation corresponds to the homoclinic bifurcation with two saddle-loops at the same saddle representing the unstable partially synchronized symmetric state.
The pitchfork bifurcation values $\teps = \teps_{\mathrm{PF}}(\tomega)$ and $\tomega = \tomega_{\mathrm{PF}}(\teps)$ can be found from Eqs.~\eqref{eq:cubic-varrho-1} and \eqref{eq:cubic-varrho-2}.
By eliminating $\varrho$ from these equations, one obtains the pitchfork bifurcation line
\begin{gather*}
  \label{eq:pitchfork}
  \tomega^4 + 2 \left(3\teps + 1\right) \tomega^2 - \left(\teps - 1\right)^3 = 0
\end{gather*}
starting at point $(\teps, \tomega) = (4, 1)$ (with $\tomega > 1$, $\teps > 4$).
Similarly to the saddle-node bifurcation, the equation for the pitchfork bifurcation is biquadratic in $\tomega$ and cubic in $\teps$ allowing analytical solution
\begin{gather*}
  \teps_{\mathrm{PF}}\f{\tomega} = \left(1 + \tomega^{2/3}\right)^2, \\
  \tomega_{\mathrm{PF}}\f{\teps} = \left(\sqrt{\teps} - 1\right)^{3/2}
\end{gather*}
with $\abs{\mathcal R} = \sqrt{1/\sqrt{\teps}-2/\teps}$.
The parametrization in terms of angle $\psi$ is as follows:
\begin{gather*}
  \tomega = \tau^3 = \tan^3 \frac{\psi}2, \\
  \teps = \left(1 + \tau^2\right)^2 = \cos^{-4} \frac{\psi}2, \\
  \abs{\mathcal R} = \frac{\sqrt{\tau^2 - 1}}{\tau^2 + 1} = \cos \frac{\psi}2 \sqrt{- \cos\psi}
\end{gather*}
with $\pi/2 < \psi < \pi$, $\tau > 1$.

\paragraph{Oscillatory regime.}
At $\tomega > 1$, $\teps = 4$, the stable oscillatory regime (limit cycle) is born from the stable asynchronous state through the degenerate Andronov–Hopf bifurcation.
The period of oscillations grows with $\teps$ and tends to infinity when approaching homoclinic (at $1 < \tomega < \tomega_{\mathrm{SNL}}$) or saddle-node infinite-period bifurcation (at $\tomega > \tomega_{\mathrm{SNL}}$) where the limit cycle disappears.
In oscillations, the global order parameter $\abs{\mathcal R}$ changes between zero and some maximum value (hitting zero and maximum twice on the period) whereas the subgroup order parameters $\mathcal Z_{1, 2}$ are never equal to zero.
The parameter values for the saddle-node infinite-period bifurcation are given in the previous paragraph.

Using the Poincare–Lindstedt method, one can obtain the asymptotic expansion for the normalized oscillation frequency $\tOmega_{\mathrm{lc}}$ near the degenerate Andronov–Hopf bifurcation, at $\teps \to 4$,
\begin{multline*}
  \frac{\tOmega_{\mathrm{lc}}}{\sqrt{\tomega^2 - 1}} = 1 - \frac{5}{8} \frac{\teps - 4}{\tomega^2 - 1} \\
  {}- \frac{32\tomega^4 + 15\tomega^2 + 12}{256 \tomega^2} \left(\frac{\teps - 4}{\tomega^2 - 1}\right)^2 + \mathcal O\f{\left(\teps - 4\right)^3}.
\end{multline*}
The maximum and period-average values of $\abs{\mathcal R}^2$ are $\teps/4 - 1 + \mathcal O\f{(\teps - 4)^2}$ and $\teps/8 - 1/2 + \mathcal O\f{(\teps - 4)^2}$, respectively.
The extremum values and period-average values of $\rho_{1, 2}^2$ are given by $(\teps/4 - 1)(1 \pm 1/\tomega) + \mathcal O\f{(\teps - 4)^2}$ and $\teps/4 - 1 + \mathcal O\f{(\teps - 4)^2}$, respectively.

For the homoclinic bifurcation at $1 < \tomega < \tomega_{\mathrm{SNL}}$, the corresponding bifurcation value $\teps = \teps_{\mathrm{HC}}(\tomega)$ can be found numerically.
This bifurcation value lies between the saddle-node and pitchfork ones, $\teps_{\mathrm{SN}}(\tomega) < \teps_{\mathrm{HC}}(\tomega) < \teps_{\mathrm{PF}}(\tomega) < 2 + 2\tomega^2$ and can be approximated as $\teps_{\mathrm{HC}}(\tomega) \approx \teps_{\mathrm{SNL}} + 2.0893 (\tomega - \tomega_{\mathrm{SNL}}) + 0.445 (\tomega - \tomega_{\mathrm{SNL}})^2 - 0.1003 (\tomega - \tomega_{\mathrm{SNL}})^3$ with absolute error not exceeding $4 \times 10^{-4}$.
\begin{figure*}{}
  \includegraphics[width=\textwidth]{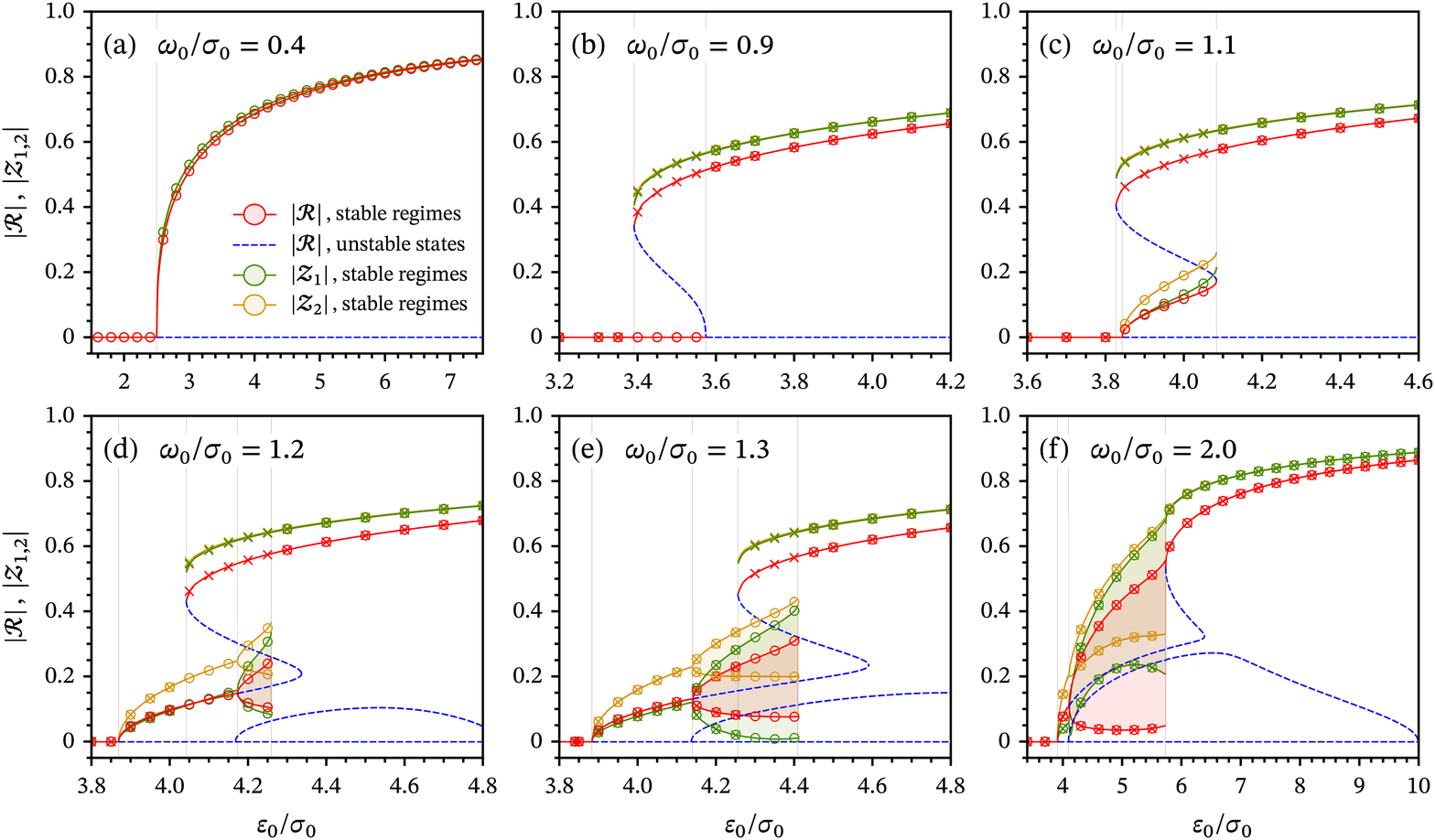}
  \caption{\label{fig:3} Synchronization branches for dependences of the global $\abs{\mathcal{R}}$ and subgroup $\abs{\mathcal{Z}_{1, 2}}$ order parameters on the normalized coupling strength $\varepsilon_0/\sigma_0$ in the slightly asymmetric case $\Dg = 0.02$, $\De = \Ds = \Da = 0$ for different normalized subgroup frequency differences $\omega_0/\sigma_0$: (a) $0.4$, (b) $0.9$, (c) $1.1$, (d) $1.2$, (e) $1.3$, and (f) $2.0$.
  For the stable stationary states, the red, green, and yellow lines or markers correspond, respectively, to the global $\abs{\mathcal{R}}$, first $\abs{\mathcal{Z}_{1}}$, and second $\abs{\mathcal{Z}_{2}}$ subgroup order parameters.
  For the periodic regimes, the solid lines and markers denote the minimal and maximal (over period) values of the order parameter and the area between them is shaded.
  The blue dashed lines indicate the value of the global order parameter $\abs{\mathcal{R}}$ in the unstable stationary states.
  The gray vertical lines mark the bifurcations of the stable regimes.
  All lines are obtained from the low-dimensional system~\eqref{eq:Z1}–\eqref{eq:R}.
  The markers show the results found from the numerical solution of the kinetic equation~\eqref{eq:kinetic} in the setup where the coupling parameter $\varepsilon_0$ gradually increases (round markers) or decreases (cross markers).}
\end{figure*}

\section{Asymmetric case with different subgroup populations}
\label{sec:4}

Now let us consider the asymmetric case with $\Dg \neq 0$, $\Dte = \Dts = \Da = 0$.
Figures~\ref{fig:1} and \ref{fig:3} summarize the results of our analysis.
Figure~\ref{fig:1} shows the bifurcation maps for different values of asymmetry, and Fig.~\ref{fig:3} presents the respective synchronization branches $\abs{\mathcal R(\teps)}, \abs{\mathcal Z_{1,2}(\teps)}$ for various $\tomega$ and slight asymmetry $\Dg = 0.02$.
Similarly to the symmetric case, the system may support incoherent and stationary partially synchronized states as well as the oscillatory regime as described below.
The main difference with the symmetric case is that there may exist a second stable partially synchronized state.
This state correspond to the one of two abovedescribed asymmetric unstable partially synchronized states in the symmetric case which may become stable due to the introduction of the asymmetry into the system.

\paragraph{Incoherent state.}
The characteristic equation of the Eqs.~\eqref{eq:normalized-Z1}–\eqref{eq:normalized-R} linearized around $\mathcal Z_{1, 2} = 0$ is
\begin{gather*}
  \left[\left(\lambda + 1\right)^2 - \frac{\teps}2 \left(\lambda + 1\right) + \tomega^2\right]^2 + \frac{\tomega^2\teps^2 \Dg^2}4 = 0,
\end{gather*}
where $\lambda$ is the characteristic variable.
The bifurcation manifold where the incoherent state loses stability can be found through the Neimark's method of D-partitions~\cite{gryazina_d-decomposition_2004}, that is, by taking purely imaginary $\lambda$ in the characteristic equation and analyzing the solvability conditions.
This gives the equation
\begin{gather}
   \label{eq:andronov-hopf}
   \left(\teps - 4\right)^2 \left(\teps - 2 - 2\tomega^2\right) + 2\tomega^2 \teps^2 \Dg^2 = 0
\end{gather}
for the bifurcation value $\teps = \teps_{\mathrm{AH}1}(\tomega, \Dg)$ of the Andronov–Hopf bifurcation with $2 \leqslant \teps_{\mathrm{AH}1}(\teps, \Dg) \leqslant 4/(1 + \abs{\Dg})$ (equalities are possible only for symmetric case or limits $\tomega \to 0, \infty$).
The respective asymptotics for the bifurcation value of the coupling $\teps$ are given in the Appendix~\ref{app:asymmetric1:ah1}.
Contrary to the symmetric case, the incoherent state always loses stability through the Andronov–Hopf bifurcation, and the bifurcation value is always lower than in the symmetric case.
As a result of the degenerate Andronov–Hopf bifurcation at $\teps = \teps_{\mathrm{AH}1}(\tomega, \Dg)$, a stable partially synchronous state is born with rotation frequency $\tOmega_s = \tOmega_{\mathrm{AH}1}(\tomega, \Dg)$ and $\psi = \psi_{\mathrm{AH}1}(\tomega, \Dg)$, where
\begin{gather*}
  \begin{multlined}
    \tOmega_{\mathrm{AH}1}\f{\tomega, \Dg} = \frac{\Dg \tomega\teps_{\mathrm{AH}1}\f{\tomega, \Dg}}{4 - \teps_{\mathrm{AH}1}\f{\tomega, \Dg}} \\
    {}= \sgn \Dg \sqrt{1 + \tomega^2 - \frac{\teps_{\mathrm{AH}1}(\tomega, \Dg)}2},
  \end{multlined} \\
  \begin{multlined}
    \psi_{\mathrm{AH}1}\f{\tomega, \Dg} = \arctan \frac{4\tomega}{4 - \teps_{\mathrm{AH}1}\f{\tomega, \Dg}} \\
    {}= \arctan\f{\tomega - \tOmega_{\mathrm{AH}1}\f{\tomega, \Dg}} \\
    {}+ \arctan\f{\tomega + \tOmega_{\mathrm{AH}1}\f{\tomega, \Dg}}.
  \end{multlined}
\end{gather*}
As follows from these formulas for small asymmetries $\abs{\Dg} \ll 1$, the normalized synchronization frequency $\tOmega_{\mathrm{AH}1}(\tomega, \Dg)$ is close to the median frequency $0$ at $\tomega \lesssim 1$, $\abs{\tOmega_{\mathrm{AH}1}(\tomega, \Dg)} \ll \tomega$, and approaches the natural frequency of stronger subgroup with growing $\tomega$ for higher $\tomega \gtrsim 1$, $\tOmega_{\mathrm{AH}1}(\tomega, \Dg) \approx \sgn \Dg \sqrt{\tomega^2 - 1}$.
The angle $\psi_{\mathrm{AH}1}\f{\tomega, \Dg}$ increases from $0$ to approximately $\pi/2$ when $\tomega$ changes from $0$ to $1$ and stays almost constant afterwards.
For large asymmtetries $\abs{\Dg} \approx 1$, the synchronization frequency is close to the natural frequency of the stronger subgroup, $\tOmega_{\mathrm{AH}1}(\tomega, \Dg) \approx \tomega \sgn \Dg$, with angle $\psi \approx \arctan 2\tomega$.

Note that, if
\begin{multline*}
  \left(\tomega^2 - 1\right)^3 - \tomega^2 \left(2\tomega^4 + 14\tomega^2 + 11\right) \Dg^2 + \tomega^4 \left(\tomega^2 + 1\right) \Dg^4 \\
  > 0,
\end{multline*}
Eq.~\eqref{eq:andronov-hopf} has two additional solutions for $\teps$ with $\teps > 4/(1 - \abs{\Dg})$ that correspond to the additional degenerate Andronov–Hopf bifurcations where the incoherent state remains unstable, but an unstable partially synchronized state is born and then disappears.
This unstable state is a result of removal of symmetry-conditioned degeneracy.
In the symmetric case, this state corresponds to the asymmetric unstable state (with frequency close to the frequency of the smaller oscillator subgroup) at lower $\teps$ and to the symmetric unstable state (born from saddle-node bifurcation) at higher $\teps$.

\paragraph{Partially synchronized states.}
To study partially synchronized states one may obtain polynomial equations for the global and local order parameters.
To this end, one may use Eqs.~\eqref{eq:normalized-Z1} and \eqref{eq:normalized-Z2} in the rotating frame (at frequency $\tOmega_s$) where $\mathcal R$ is constant and positive.
By equating the right-hand sides of those equations to zero and taking the condition~\eqref{eq:normalized-R}, one obtains 3 complex algebraic equations for 2 complex ($\mathcal Z_{1, 2}$) and 2 real ($\mathcal R$, $\tOmega_s$) variables.
Then one may eliminate all variables except $\mathcal R$ and obtain a single polynomial equation $P(\varrho, \teps, \tomega^2, \Dg^2) = 0$ of the ninth degree with respect to $\varrho = \teps\abs{\mathcal R}^2 = \teps \mathcal R^2$.
One can perform this elimination of variables in two steps.
First, $\mathcal Z_{2}$ can be expressed from Eq.~\eqref{eq:normalized-R} and substituted in polynomials for $\Dd{\mathcal Z_{1, 2}}/\Dd{\tlt} - \I \tOmega_s\mathcal Z_{1, 2}$ obtained from Eqs.~\eqref{eq:normalized-Z1} and \eqref{eq:normalized-Z2}.
After that one can eliminate $\mathcal Z_1$ by finding the resultant of these polynomials.
Second, one computes the resultant of the real and imaginary parts of the previous resultant with respect to $\tOmega_s$, reduces it by $\varrho$, and finds $P(\varrho, \teps, \tomega^2, \Dg^2)$.
Unfortunately, the polynomial equation $P(\varrho, \teps, \tomega^2, \Dg^2) = 0$ is already too cumbersome to study analytically in general case, but it facilitates the numerical studies and can be used in some limiting cases to obtain analytical results (which are mostly presented in Appendix~\ref{app:asymmetric1}).

As the numerical study showed, at $\teps > \teps_{\mathrm{AH}1}(\tomega, \Dg)$ there may be one, two, or three solutions of $P(\varrho, \teps, \tomega^2, \Dg^2) = 0$ with $0 < \varrho < \teps$ that correspond to the partially synchronized states.
By equating the discriminant of $P$ with respect to $\varrho$ to zero, one obtains the conditions where the number of the partially synchronized states may change due to the saddle-node bifurcations.
There are three solutions only in the region bounded by the two saddle-node bifurcation lines $\teps = \teps_{\mathrm{SN}1}(\tomega, \Dg)$, $\tomega > \tomega_{\mathrm{SN}1}(\Dg)$ and $\teps = \teps_{\mathrm{SN}2}(\tomega, \Dg)$, $\tomega > \tomega_{\mathrm{SN}2}(\Dg)$; there exists only one solution outside this region; and two distinct solutions happen at the saddle-node bifurcations.
One of the saddle-node bifurcations $\teps = \teps_{\mathrm{SN}1}(\tomega, \Dg)$ corresponds in the symmetric case to the saddle-node bifurcation $\teps = \teps_{\mathrm{SN}}(\tomega)$, while the other one $\teps = \teps_{\mathrm{SN}2}(\tomega, \Dg)$ appears from the pitchfork bifurcation $\teps = \teps_{\mathrm{PF}}(\tomega)$ due to the degeneracy removal.
To be more precise, the bifurcation at $\teps = \teps_{\mathrm{SN}1}(\tomega, \Dg)$ can be either of saddle-node or saddle-node infinite-period type, similarly to how it is in the symmetric case.
The asymptotic behavior of the saddle-node bifurcation lines is described in Appendix~\ref{app:asymmetric1:sn}.

\begin{figure*}{}
  \includegraphics[width=\textwidth]{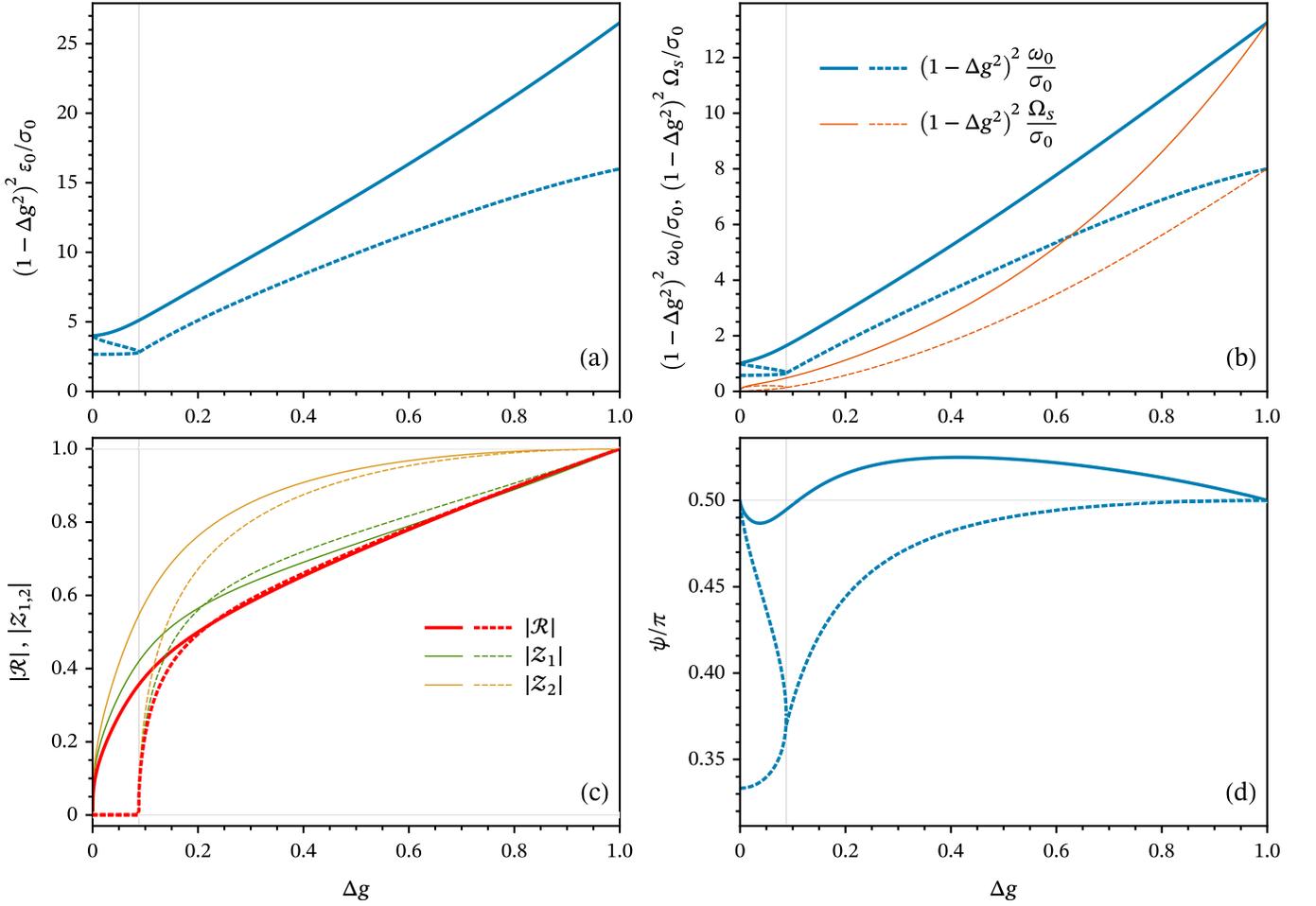}
  \caption{\label{fig:4} The parameters and the dynamical variables at the codimension-2 bifurcations as function of the asymmetry parameters $\Dg$.
  Solid lines correspond to the coherent Bogdanov–Takens bifurcation, and the dashed lines indicate the incoherent Bogdanov–Takens bifurcations at $\Dg < \Dg_c \approx 0.088$ and the cusp bifurcation at $\Dg > \Dg_c$.
  The position of the codimension-3 bifurcation at $\Dg = \Dg_c$ is marked by the gray vertical gridlines.
  (a) The bifurcation values of the normalized coupling constant $(1 - \Dg^2)^2 \varepsilon_0/\sigma_0$.
  (b) The bifurcation values of the normalized subgroup frequency difference $(1 - \Dg^2)^2 \omega_0/\sigma_0$ and the normalized synchronization frequency $(1 - \Dg^2)^2 \Omega_s/\sigma_0$ of the respective stationary state.
  (c) The global $\abs{\mathcal R}$ and subgroup $\abs{\mathcal Z_{1,2}}$ order parameters of the stationary state.
  (d) The phase shift $\psi$ between the subgroup order parameters.}
\end{figure*}

At small asymmetries, $\abs{\Dg} < \Dg_{c} \approx 0.0878$, the two saddle-node bifurcation lines start at two distinct Bogdanov–Takens points
\begin{multline*}
  \left(\teps, \tomega\right) = \left(\teps_{\mathrm{AH}1}\f{\tomega_{\mathrm{SN}1, 2}(\Dg), \Dg}, \tomega_{\mathrm{SN}1, 2}\f{\Dg}\right) \\
  {}= \left(\teps_{\mathrm{SN}1, 2}\f{\tomega_{\mathrm{SN}1, 2}(\Dg), \Dg}, \tomega_{\mathrm{SN}1, 2}\f{\Dg}\right)
\end{multline*}
on the Andronov–Hopf bifurcation line $\teps = \teps_{\mathrm{AH}1}(\tomega, \Dg)$ like in Figs.~\ref{fig:1}(b, d).
At larger asymmetries, $\abs{\Dg} > \Dg_{c}$ both saddle-node bifurcation lines start at the cusp point $(\teps, \tomega) = (\teps_{\mathrm{C}}(\Dg), \tomega_{\mathrm{C}}(\Dg))$ so that $\tomega_{\mathrm{SN}1}(\Dg) = \tomega_{\mathrm{SN}2}(\Dg) = \tomega_{\mathrm{C}}(\Dg)$, $\teps_{\mathrm{SN}1}(\tomega_{\mathrm{SN}1}(\Dg), \Dg) = \teps_{\mathrm{SN}2}(\tomega_{\mathrm{SN}2}(\Dg), \Dg) = \teps_{\mathrm{C}}(\Dg)$ like in Figs.~\ref{fig:1}(c, e).
The asymptotic expression for the positions of those Bogdanov–Takens points and the cusp point are given in Appendix~\ref{app:asymmetric1:ibt}.
The critical asymmetry value $\Dg_{c}$ is an algebraic number that can be expressed in radicals.
Since this critical value corresponds to the cusp point lying on the Andronov–Hopf bifurcation line, one may find $\Dg_{c}$ by eliminating $\teps$ and $\tomega$ from Eq.~\eqref{eq:andronov-hopf}, $P'(0, \teps, \tomega^2, \Dg^2) = 0$, and $P''(0, \teps, \tomega^2, \Dg^2) = 0$ (where prime denotes the derivative with respect to the first argument).
As a result, one obtains the biquartic equation for $\Dg_{c}$,
\begin{gather*}
  162 \Dg_{c}^8 - 5850 \Dg_{c}^6 - 18631 \Dg_{c}^4 - 3356 \Dg_{c}^2 + 27 = 0,
\end{gather*}
$\Dg_{c} \approx 0.08783$, $\tomega_{\mathrm{C}}(\Dg_{c}) \approx 0.6794$, $\teps_{\mathrm{C}}(\Dg_{c}) \approx 2.8766$.
The positions of the cusp point (at $\Dg > \Dg_c$) and incoherent Bogdanov–Takens bifurcations (at $\Dg < \Dg_c$) together with the respective values of dynamical variables are shown in Fig.~\ref{fig:4} as functions of the asymmetry parameters $\Dg$.

We examined the stability of the partially synchronized states numerically by computing eigenvalues of the linearized system~\eqref{eq:rho1}–\eqref{eq:psi} and analytically.
The analytical study is based on the computing and factorizing the resultant of $P(\varrho, \teps, \tomega^2, \Dg^2)$ and another polynomial which changes sign when partially synchronized state loses stability (this polynomial is obtained using Neimark's D-partition method).
In the region between two saddle-node bifurcations, where there are 3 partially synchronized states, the state with the largest order-parameter is always stable and the state with second-largest one is always unstable.
As for the third state and the only state in the outside region, it can be as stable, as unstable.
This stability loss occurs through the another (‘coherent’) Andronov–Hopf bifurcation giving birth to the oscillatory regime.
The bifurcation line corresponding to this Andronov–Hopf bifurcation starts at the Bogdanov–Takens point lying on the saddle-node curve $\teps = \teps_{\mathrm{SN}2}(\tomega, \Dg)$ and approaches the asymptote $\teps = 4/(1 - \abs{\Dg})$ at $\tomega \to \infty$.
Appendix~\ref{app:asymmetric1:ah2} gives the respective asymptotics for this bifurcation value $\teps = \teps_{\mathrm{AH}2}(\tomega, \Dg)$ and the parameters of the partially synchronized state at the stability boundary, and Appendix~\ref{app:asymmetric1:cbt} does so for the coherent Bogdanov–Takens point.
The values of parameters and dynamical variables for the coherent Bogdanov–Takens bifurcation values are plotted versus $\Dg$ in the Fig.~\ref{fig:4}.

Note that, strangely enough, the straight line $\teps = 4/(1 - \abs{\Dg})$ is also an asymptote for the Andronov–Hopf bifurcation of the incoherent state given by Eq.~\eqref{eq:andronov-hopf} (where no stable regime participates), though these two Andronov–Hopf bifurcations happen at different points in the phase space and are apparently unrelated.
At small asymmetries, the two bifurcation curves are so close that it is hard to distinguish them apart in the Figs.~\ref{fig:1}(b, d) outside the immediate vicinity of the saddle-node curve $\teps = \teps_{\mathrm{SN}2}(\tomega, \Dg)$ and the Bogdanov–Takens point.
Note also that, at large asymmetries, the bifurcation line can make a U-like turn in $(\teps, \tomega)$ plane, so there may be two coherent Andronov–Hopd bifurcations for low enough fixed $\tomega$, as seen in Fig.~\ref{fig:1}(c, e): the coherent state first loses stability and then regains it back with growing $\teps$. 

One can also obtain analytical formulas for the order parameters of the partially synchronized state in the limiting cases of small and large couplings, $\teps \to \teps_{\mathrm{AH}2}(\tomega, \Dg), \infty$, and for small and large frequency differences, $\tomega \to 0, \infty$, as presented in Appendices~\ref{app:asymmetric1:couplings} and \ref{app:asymmetric1:omega}, respectively.
As seen, in the limit $\teps \to \infty$, the synchronization becomes global, $\abs{\mathcal R} \simeq 1$, with synchronization frequency approaching the mean natural frequency, $\tOmega_s \simeq \Dg \tomega$, and the angle $\psi$ tending to zero, $\psi \simeq 2\tomega/\teps$.

Let us discuss here the limit $\tomega \to \infty$ in more detail.
By applying this limit to the equation $P(\varrho, \teps, \tomega, \Dg) = 0$, one obtains two solutions
\begin{gather*}
  \varrho_{\pm} \simeq \frac{\left(1 \pm \abs{\Dg}\right)^2}4 \left(\teps - \frac{4}{1 \pm \abs{\Dg}}\right).
\end{gather*}
Solutions $\varrho_{\pm}$ are born from the incoherent state through the Andronov–Hopf bifurcations at $\teps \simeq 4/(1 \pm \abs{\Dg})$.
The analytical stability analysis in the limit $\tomega \to \infty$ shows that the solution with $\varrho = \varrho_{-}$ is always unstable, whereas the one with $\varrho = \varrho_{+}$ is stable at
\begin{gather}
  \label{eq:infinite-omega-condition}
  \frac{4}{1 + \abs{\Dg}} < \teps < \frac{4}{1 - \abs{\Dg}}
\end{gather}
with the syncrhonization frequency equal to the natural frequency of the stronger subgroup, $\tOmega_s \simeq \sgn \Dg\, \tomega$, and the order parameter
\begin{gather}
  \label{eq:infinite-omega-R}
  \mathcal R = \sqrt{\frac{\varrho_{+}}{\teps}} \simeq \frac{1 + \abs{\Dg}}2\sqrt{1 - \frac{4}{\left(1 + \abs{\Dg}\right)\teps}}
\end{gather}
that increases with $\teps$ from $0$ to $\mathcal R_m = \sqrt{\abs{\Dg} (1 + \abs{\Dg})/2}$, after which the Andronov–Hopf bifurcation occurs and the stationary state loses stability.
At $\teps = 4$, the order parameter is $\mathcal R = \mathcal R_m/\sqrt{2}$.
This expression for $\mathcal R_m$ gives actually the lower estimate for the value of the order parameter at the stability boundary: this quantity grows slowly with decreasing $\tomega$ and achieves greater values at the Bogdanov–Takens point (around $\sqrt[3]{\abs{\Dg}}/2^{5/6}$ for small asymmetries $\abs{\Dg}$), as seen from formulas in Appendix~\ref{app:asymmetric1:ah2} and~\ref{app:asymmetric1:cbt}.
Note that even in the limit $\tomega \to \infty$, the stationary order parameter remains finite, so the two Andronov–Hopf bifurcation indeed happen simultaneously at two different points of the phase space.
Also note that at $\tomega \to \infty$, there is no bistability and the stationary state~\eqref{eq:infinite-omega-R} is the only stable regime if conditions \eqref{eq:infinite-omega-condition} are fulfilled.

\begin{figure*}{}
  \includegraphics[width=\textwidth]{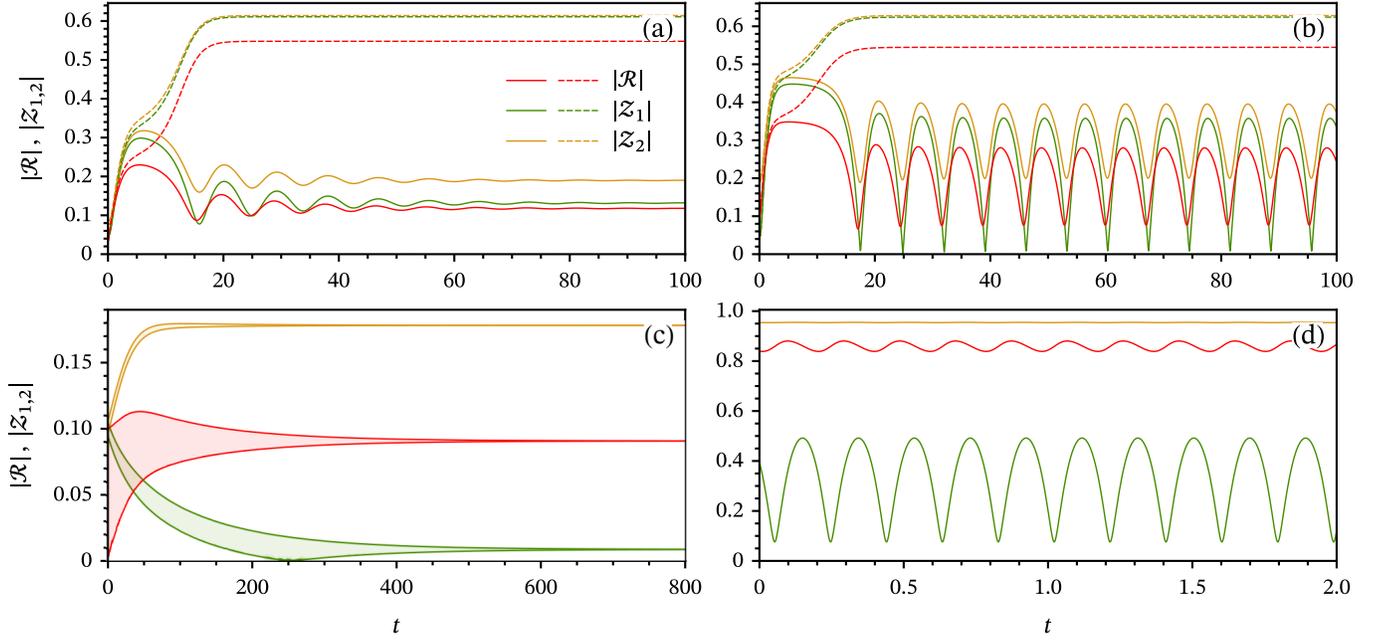}
  \caption{\label{fig:5} The time dependences of the global $\mathcal R$ (red lines) and subgroup $\mathcal Z_{1, 2}$ (green and yellow lines) order parameters for different parameters and initial conditions: (a) $\Dg = 0.02$, $\varepsilon_0 = 4$, $\omega_0 = 1.1$, $\mathcal Z_1(0) = 0.04 + 0.04\I$, $\mathcal Z_2(0) = -0.08\I$ (solid lines), $-0.09\I$ (dashed lines), (b) $\Dg = 0.02$, $\varepsilon_0 = 4.35$, $\omega_0 = 1.3$, $\mathcal Z_1(0) = 0.04 + 0.04\I$, $\mathcal Z_2(0) = -0.19\I$ (solid lines), $-0.2\I$ (dashed lines), (c) $\Dg = 0.02$, $\varepsilon_0 = 4.05$, $\omega_0 = 10$, $\mathcal Z_1(0) = \mathcal Z_2(0) = 0.1$, and (d) $\Dg = 0.8$, $\varepsilon_0 = 25$, $\omega_0 = 20$, $\mathcal Z_1(0) = -0.0982 - 0.192\I$, $\mathcal Z_2(0) = 0.9 + 0.3168\I$.
  In all panels, $\sigma_0 = 1$, $\De = \Ds = \Da = 0$.
  In panel (c), the oscillations are too quick to be distinguishable with the chosen scale, and the lines with shading between them represent the minimum and maximum envelopes.}
\end{figure*}

Thus, the introduction of the asymmetry results in stabilization of initially (in the symmetric case) unstable partially synchronized state.
Such states in the asymmetric case were previously found and described in~\cite{acebron_asymptotic_1998, acebron_breaking_1998} for the problem with the Gaussian noise.
Here, the specifics of the Cauchy noise allowed us to describe analytically the stability boundary for the partially synchronized state as well as upper limits for the order parameter.
The non-empty region of stability exists at any $\tomega$ (including infinitely large $\tomega$) and any non-zero $\Dg$, and the maximal order parameter remains finite.
Moreover, at small $\Dg$ the maximal order parameter grows quickly as $\sqrt{\abs{\Dg}/2}$, and even slight asymmetry can lead to noticeable synchronization.
For $\Dg = 0.02$, when two subgroups contain $51\%$ and $49\%$ of all oscillators, the order parameter is already on the level of $0.1$.
This can be seen both from the obtained analytical formula for $\mathcal R_m$ and numerical results in Figs.~\ref{fig:3}(c–f).
This can be also observed in Fig.~\ref{fig:5} where the time dependences of the order parameters are shown in few exemplary cases.
In particular, Fig.~\ref{fig:5}(a, c) show examples of how initial conditions may evolve to the state with the global order parameter around $\sqrt{{\Dg}/2} = 0.1$.

At smaller $\tomega$, this asymmetry-induced stabilization may lead to bistability between two partially synchronized states that can be seen in Figs.~\ref{fig:1}(c–e), \ref{fig:3}(c, d) and \ref{fig:5}(a).
This kind of bistability is impossible in the symmetric case.
Note that the two subgroups have significantly different order parameters $\rho_{1, 2} = \abs{\mathcal Z_{1, 2}}$ in this newly stabilized state with lower synchronization level whereas the two subgroups have almost equal order parameters in more symmetric state with higher global order parameter.
These states also differ by the synchronization frequency $\tOmega_s$.
The frequency of the lower state is noticeably shifted towards the natural frequency of the subgroup with higher population while the higher state is synchronized at approximately the mean frequency.
All this indicates the high sensitivity even to a small asymmetry. 

Fig.~\ref{fig:5}(b) illustrates another kind of bistability when the oscillatory regime coexists with a partially synchronized state.
This type of bistability could be observed in the symmetric case, however, in asymmetric case, the oscillatory regime may be also significantly asymmetric even for small $\Dg$.

Let us also note that the Andronov–Hopf bifurcation line (where this lower state loses stability) may approach the Bogdanov–Takens point from different sides depending on the asymmetry degree $\Dg$, compare Figs.~\ref{fig:1}(b, d) and \ref{fig:1}(c, e).
At higher asymmetries, the Andronov–Hopf bifurcation line goes out of the Bogdanov–Takens point towards decreasing $\teps$ and $\tomega$, so there are values of $\tomega$ to which correspond two bifurcation values of $\teps$: with increasing $\teps$ the stationary state first loses stability and then reclaims it through the Andronov–Hopf bifurcations.

The equation $P(\varrho, \teps, \tomega^2, \Dg^2) = 0$ can also be used for examining the noiseless case. 
To this end, one may consider the limit $\sigma_0 \to 0$ or, what is the same, the limit $\teps, \tomega \to \infty$ with $\teps/\tomega = \varepsilon_0/\omega_0$ fixed.
For $\varepsilon_0/\omega_0 < 2$, one finds that there is no stable states in the noiseless case and that the only stable regime possible is the oscillatory one.
At $\varepsilon_0 = 2\omega_0$, the saddle-node infinite-period bifurcation happens, and two partially synchronized states are born, stable and unstable.
At $\varepsilon_0/\omega_0 > 2$, this stable state has the order parameter
\begin{gather*}
  \abs{\mathcal R} = \sqrt{\frac{1 + \Dg^2}2 + \frac{1 - \Dg^2}2 \sqrt{1 - \frac{4\omega_0^2}{\varepsilon_0^2}}} = \sqrt{\frac{\xi^2 + \Dg^2}{\xi^2 + 1}}
\end{gather*}
with $\Omega_s = \Dg \omega_0$, $\rho_{1, 2} = 1$ and
\begin{gather*}
  \psi = 2\arcsin \sqrt{\frac 12 \left(1 - \sqrt{1 - \frac{4\omega_0^2}{\varepsilon_0^2}}\right)} = 2 \arctan \frac 1{\xi},
\end{gather*}
where
\begin{gather*}
  \xi = \frac{\varepsilon_0}{2\omega_0} + \sqrt{\frac{\varepsilon_0^2}{4\omega_0^2} - 1} = \exp \mathop{\mathrm{arccosh}} \frac{\varepsilon_0}{2\omega_0} > 1.
\end{gather*}
The asymptotic corrections to the above noiseless solution in the case of a small, but finite noise are as follows:
\begin{gather*}
  \begin{multlined}
    \abs{\mathcal R} = \sqrt{\frac{\xi^2 + \Dg^2}{\xi^2 + 1}} \twolinebrackets{\{}{\}}{1 - \frac{\xi}{\xi^4 - \Dg^2} {\left[\xi^2 + 1 - 2\Dg^2\vphantom{\xi^2 + 1 - 2\Dg^2{}+ \frac{1 - \Dg^2}{1 + \Dg^2} \left(\frac{1 - \Dg^2}{\xi^2 - 1} - \Dg^2 \frac{1 + 3\Dg^2}{\xi^2 + \Dg^2}\right)}\right.} }%
    {\left.\vphantom{\xi^2 + 1 - 2\Dg^2{}+ \frac{1 - \Dg^2}{1 + \Dg^2} \left(\frac{1 - \Dg^2}{\xi^2 - 1} - \Dg^2 \frac{1 + 3\Dg^2}{\xi^2 + \Dg^2}\right)}{}+ \frac{1 - \Dg^2}{1 + \Dg^2} \left(\frac{1 - \Dg^2}{\xi^2 - 1} - \Dg^2 \frac{1 + 3\Dg^2}{\xi^2 + \Dg^2}\right)\right]\frac{\sigma_0}{\omega_0} + \mathcal O\f{\sigma_0^2}},
  \end{multlined} \\
  \Omega_s = \Dg\omega_0 + \frac{\xi \Dg \left(1 - \Dg^2\right)}{\xi^4 - \Dg^2} \sigma_0 + \mathcal O\f{\sigma_0^2}, \\
  \rho_{1, 2} = 1 - \frac{\xi}{\xi^2 \mp \Dg} \frac{\sigma_0}{\omega_0} + \mathcal O\f{\sigma_0^2}, \\
  \psi = 2 \arctan \frac 1{\xi} + \frac{2 \xi^2 \left(\xi^2 - \Dg^2\right)}{\left(\xi^2 - 1\right) \left(\xi^4 - \Dg^2\right)} \frac{\sigma_0}{\omega_0} + \mathcal O\f{\sigma_0^2}.
\end{gather*}

Another curious remark concerns the specific value of the angle $\psi = \pi/2$ in partially synchronized states.
The asymptotics presented in Appendix~\ref{app:asymmetric1} show that this value occurs quite ‘often’: for the large $\tomega$, for the ‘coherent’ Andronov–Hopf bifurcation, for the ‘coherent’ Bogdanov–Takens points, for the saddle-node bifurcation at large $\tomega$, the value of $\psi$ in the noticeably asymmetric stable state is also close to $\pi/2$ be at large $\tomega$ or $\tomega \approx 1$.
In particular, for the partially synchronized state at the Bogdanov–Takens point, $-0.042 < \psi - \pi/2 < 0.079$ for any asymmetry $\Dg$ (see Fig.~\ref{fig:4}d).
Equation~\eqref{eq:psi} for angular velocity may give a hint of why this value of $\pi/2$ is so important.
For $\Da = \Ds = \De = 0$, this equation transforms into
\begin{gather*}
  \frac{\Dd{\psi}}{\Dd{\tlt}} = 2\tomega - \frac{\teps}4 \left[\left(1 - \Dg\right) \frac{\rho_1}{\rho_2} + \left(1 + \Dg\right) \frac{\rho_1}{\rho_2} + 2\rho_1 \rho_2\right] \sin\psi.
\end{gather*}
As seen, to counteract strong rotating momentum $2\tomega$, a strong binding force is needed which is proportional to coupling strength, $\sin\psi$, and some positive function of subgroup order parameters $\rho_{1, 2}$.
With $\teps$ and $\rho_{1, 2}$ fixed, the strongest binding force achieved at $\psi = \pi/2$.
Therefore, one may expect that large $\tomega$ will require $\psi$ around $\pi/2$ at the boundaries of existence of the partially synchronized states (formed by saddle-node bifurcations and Andronov—Hopf bifurcation of the incoherent state; note, by the way, that the position of the first saddle-node bifurcation depends on the asymmetry very weakly at large $\tomega$, $\teps_{\mathrm{SN}1}(\tomega, \Dg) \approx 2\tomega + 2$).
This is, of course, a quite unrigorous reasoning and does not explain why the loss of stability also takes place at $\psi \approx \pi/2$ even for not so large $\tomega$.
However, we may speculate further that the Andronov–Hopf bifurcation of the coherent state at small asymmetries is close to the Andronov–Hopf bifurcation of the incoherent state, so naturally the value of $\psi$ should be close too.
And for larger asymmetries, the Andronov–Hopf bifurcation of the coherent state may happen only at large enough values $\tomega$, $\tomega \gtrsim \teps/2 - 1$.

\paragraph{Oscillatory regime.}
At large enough frequency differences $\tomega$, the oscillatory regime that is born from the Andronov–Hopf bifurcation may disappear through homoclinic or the saddle-node infinite-period bifurcation similarly to how it is in the symmetric case.
The homoclinic bifurcation line starts at the abovementioned Bogdanov–Takens point and approaches the saddle-node-loop point at the saddle-node bifurcation line $\teps = \teps_{\mathrm{SN}1}(\tomega, \Dg)$.
Beyond the saddle-node-loop point the saddle-node corresponds to the saddle-node infinite-period bifurcation.

Note that homoclinic bifurcation may approach the Bogdanov–Takens point from different sides depending on the asymmetry degree $\Dg$ same as the Andronov–Hopf bifurcation line as described above and shown in Figs.~\ref{fig:1}(b, d) and \ref{fig:1}(c, e).
As seen in Figs.~\ref{fig:1}(c, e) for higher $\abs{\Dg}$, the homoclinic bifurcation line may go out of the Bogdanov–Takens point towards decreasing $\teps$ and $\tomega$ so that the oscillatory regime may first disappear due to the homoclinic bifurcation and then appear again to finally go away in the Andronov–Hopf bifurcation with increasing $\teps$ at fixed $\tomega$.
An example of such behavior can be seen in \ref{fig:1}(c, e) for $\Dg = 0.12$ and $\tomega = 2$ where 7 bifurcations sequentially happen when $\teps$ increases: (i) the stable incoherent state is replaced by the stable significantly asymmetric partially synchronized state in the degenerate Andronov–Hopf bifurcation; (ii) the stable partially synchronized state changes into the stable oscillatory regime in the Andronov–Hopf bifurcation; (iii) the stable highly symmetric partially synchronized state is born through saddle-node bifurcation and provides bistability between oscillatory regime and partially synchronized state; (iv) the oscillatory regime disappears into homoclinic bifurcation so the only partially synchronized state remains; (v) the second homoclinic bifurcation returns the oscillatory regime and bistability; (vi) the Andronov–Hopf bifurcation transforms the oscillatory regime into the second partially synchronized state so that the bistability of two partially synchronized states takes place; and (vii) the lower partially synchronized state disappears in saddle-node bifurcation, and only one partially synchronized state remains.

It is also worth noting that the oscillatory regime takes place for any asymmetry level $\Dg$ unless one of the subgroups completely disappears and the bimodal frequency distribution turns into the unimodal one for $\abs{\Dg} = 1$.
With $\abs{\Dg}$ growing and approaching $1$, the region of oscillatory regime in the parameter plane $(\teps, \tomega)$ is shifted towards higher values of $\tomega$, but does not completely disappear.
In other words, the limit $\abs{\Dg} \to 1$ is not uniform with respect to $\tomega$, and no matter how small is the fraction of the oscillator population is in the second subgroup, there exists an interval of $\teps$ for large enough $\tomega$ where there is no stable partially synchronized state and only stable regime is oscillatory (though the oscillation amplitude for the larger subgroup can be quite small).
An example is shown in Fig.~\ref{fig:5}(d) where the time dependences of $\abs{\mathcal R}$ and $\abs{\mathcal Z_{1, 2}}$ are shown for $\Dg = 0.8$, $\teps = 25$, and $\tomega = 20$.

The above conclusions are confirmed by the direct simulations of the kinetic equation~\eqref{eq:kinetic}.
The performed calculations showed that the Ott–Antonsen manifold that we have studied in the detail is indeed attracting, and all the obtained results remain in the high-dimensional model.

\section{Other types of asymmetry}
\label{sec:5}

The results obtained for non-zero $\Dg$ are qualitatively reproduced for the other types of asymmetry.
In particular, we compared the considered case ($\Dg \neq 0$, $\Dte = \Dts = \Da = 0$) with three other cases: (i) unequal coupling constants ($\Dte \neq 0$, $\Dg = \Dts = \Da = 0$), (ii) unequal noise levels ($\Dts \neq 0$, $\Dg = \Dte = \Da = 0$), and (iii) the non-zero phase delay in the Kuramoto–Sakaguchi system ($\Da \neq 0$, $\Dg = \Dte = \Dts = 0$).
By comparing the condition for the Andronov–Hopf bifurcation in these 3 cases,
\begin{gather*}
   \left(\teps - 4\right)^2 \left(\teps - 2 - 2\tomega^2\right) + 2\tomega^2 \teps^2 \Dte^2 = 0, \\
   \left(\teps - 4\right)^2 \left(\teps - 2 - 2\tomega^2 + 2\Dts^2\right) + 32 \tomega^2 \Dts^2 = 0, \\
  \begin{multlined}
   \left(\teps\cos\Da - 4\right)^2 \left(\teps\cos\Da - 2 - 2\tomega^2 \cos^2\Da\right) \\
   + 8\left(\teps\cos\Da - 2\right)^2 \sin^2\Da = 0,
  \end{multlined}
\end{gather*}
with Eq.~\eqref{eq:andronov-hopf}, one may deduce the approximate conversion substitutions for different types of asymmetry that leave major features of the bifurcation map approximately the same.
For unequal coupling constants, one obtains $\Dg \mapsto \Dte$.
For unequal noise levels, the substitutions are $\tomega \mapsto \sqrt{\tomega^2 - \Dts^2}$, $\Dg \mapsto -4\Dts \tomega/\sqrt{\tomega^2 - \Dts^2}$.
For the Kuramoto–Sakaguchi system, the replacement are $\teps \mapsto \teps \cos\Da$, $\tomega \mapsto \tomega \cos\Da$, $\Dg \mapsto -2(\teps \cos\Da - 2) \sin\Da / \tomega\teps\cos^2\Da$.

One may also deduce conditions where the different kind of asymmetries compensate each other, so that the Andronov–Hopf bifurcation of the incoherent state becomes degenerate at high $\tomega$.
For example, the relative difference $\De$ in coupling strengths, the relative difference in noise levels $\Ds$, and the phase delay $\Da$ can be compensated by population asymmetries $\Dg = -\De$, $\Dg = \Ds$, and $\Dg = \tan(\Da)/\tomega$, respectively.
Interestingly, that this compensation also transforms the Andronov–Hopf bifurcation into the degenerate transcritical bifurcation with $\tOmega_s = 0$ at lower $\tomega$ for $\De$ and $\Da$; however, for $\Ds$, the bifurcation at lower $\tomega$ becomes the non-degenerate Andronov–Hopf bifurcation with $\tOmega_s \neq 0$, and the rotational compensation with $\tOmega_s = 0$ occurs at another compensation value $\Dg = 2\Ds/(\tomega^2 + \Ds^2 + 1)$.

Let us also give equations for critical asymmetry values at which two ‘incoherent’ Bogdanov–Takens points merge into the cusp point; situations like in Fig.~\ref{fig:3} are possible only for asymmetry below these critical values.
For asymmetric couplings one has
\begin{gather*}
  27 \De_{c}^8 - 1260 \De_{c}^6 + 2466 \De_{c}^4 - 5356 \Dg_{c}^2 + 27 = 0
\end{gather*}
with $\De_{c} \approx 0.07108$, $\tomega_{\mathrm{C}}(\De_{c}) \approx 0.6883$, $\teps_{\mathrm{C}}(\Dg_{c}) \approx 2.9131$.
For asymmetric noise, the equation is biquadratic,
\begin{gather*}
  200 \Ds_{c}^4 + 261 \Ds_{c}^2 - 1 = 0,
\end{gather*}
with $\Ds_{c} = \sqrt{2/(261 + 41^{3/2})} \approx 0.06181$, $\tomega_{\mathrm{C}}(\Ds_{c}) = \sqrt{(53 + 7\sqrt{41})/200} \approx 0.6994$, $\teps_{\mathrm{C}}(\Ds_{c}) = (21 - \sqrt{41})/5 \approx 2.9194$.
For the Kuramoto–Sakaguchi model with phase delay, one obtains
\begin{gather*}
  4096 \sin^2\Da_c \cos^6\Da_c - 27 = 0 \\
  \text{ or }8 \sin\f{4\Da_c} + 16 \sin\f{2\Da_c} - 3\sqrt{3} = 0
\end{gather*}
with $\Da_c \approx 0.08211$,  $\tomega_{\mathrm{C}}(\Da_{c}) \approx 0.6880$, $\teps_{\mathrm{C}}(\Da_{c}) \approx 2.9135$.
One can see that not only the bifurcation picture is qualitatively the same, but the numerical values are indeed quite close for different kind of asymmetries.

\section{Conclusions}
\label{sec:6}

In conclusions, we have studied the thermodynamically large Kuramoto systems with asymmetric bimodal frequency distribution and the Cauchy noise.
To this end, the bifurcations and dynamical regimes were examined analytically and numerically in the respective low-dimensional system~\eqref{eq:Z1}–\eqref{eq:R} obtained through the Ott–Antonsen ansatz.
The obtained results were confirmed by direct numerical simulation of high-dimensional kinetic equations~\eqref{eq:kinetic}.

The introduction of the asymmetry affects drastically the dynamics around the bifurcations of the symmetric system.
In particular, the asymmetry causes the stabilization of the asymmetric stationary partially synchronized state where slightly larger oscillator subgroup is dominant with relatively high subgroup order parameter (proportional to the square root of the asymmetry degree).
This kind of stabilization takes place for arbitrarily large frequency difference between two distribution modes, and, even for infinitely large frequency difference, there exists a range of moderate coupling strengths where the only stable regime is this partially synchronized state.
For moderate frequency difference, this asymmetry-induced stabilization effect results in the new type of bistability with two stationary partially synchronized states coexisting: (i) with larger global order parameter and synchronization parity between two subgroups and (ii) with lower global order parameter and significant difference between subgroup order parameters.
The asymmetry also results in shrinking the parameter region where the bistability is possible between incoherent and partially synchronized states.
This region disappears completely at the critical asymmetry value which we have found analytically for four asymmetry types: population disparity between subgroup, different subgroup coupling strength, different noise levels, and phase delay.

Our studies reveal that even small advantage or disparity between oscillator subgroups may be decisive for the global dynamics.
This should have consequences in various application areas when there are competing synchronization forcings: for designing robust communication networks, controlling complex dynamical systems, as well as in neuroscience and for analyzing ecological systems when there are competing cells and organism subpopulations.

\begin{acknowledgments}
  The authors are grateful to A. Pikovsky and M.I. Bolotov for fruitful discussions. This work was supported by the Russian Science Foundation under project No. 22-12-00348 (analysis of the asymmetric case). GVO was supported by the Ministry of Science and Higher Education of the Russian Federation under project No. 0729-2020-0036 (analysis of the symmetric case).
\end{acknowledgments}

\appendix*
\section{Asymptotic behavior of bifurcation locations and the parameters of the partially synchronized states}
\label{app}

\subsection{Symmetric case}
\label{app:symmetric}

\subsubsection{Saddle-node and saddle-node infinite-period bifurcations}
\label{app:symmetric:sn}

The asymptotics for the position of the saddle-node or saddle-node infinite-period bifurcation and the parameters of the respective degenerate partially synchronized states are given by the following equalities:
\begin{gather*}
  \begin{multlined}
    \tomega_{\mathrm{SN}}\f{\teps} = \frac{1}{\sqrt{3}} + \frac{\sqrt{3}}{4} \left(\teps - \frac 83\right) + \frac{3\sqrt{3}}{256} \left(\teps - \frac 83\right)^2 \\
    {}+ \mathcal O\f{\left(\teps - \frac 83\right)^3},
  \end{multlined} \\
  \begin{multlined}
    \tau = \frac{1}{\sqrt{3}} + \frac{\sqrt{3}}{16} \left(\teps - \frac 83\right) - \frac{39\sqrt{3}}{2048} \left(\teps - \frac 83\right)^2 \\
    {}+ \mathcal O\f{\left(\teps - \frac 83\right)^3},
  \end{multlined} \\
  \begin{multlined}
    \psi = \frac{\pi}{3} + \frac{3\sqrt{3}}{32} \left(\teps - \frac 83\right) - \frac{135\sqrt{3}}{4096} \left(\teps - \frac 83\right)^2 \\
    {}+ \mathcal O\f{\left(\teps - \frac 83\right)^3},
  \end{multlined} \\
  \varrho = \frac{9}{16} \left(\teps - \frac 83\right) - \frac{27}{2048} \left(\teps - \frac 83\right)^2 + \mathcal O\f{\left(\teps - \frac 83\right)^3}, \\
  \rho = \frac{3}{4\sqrt{2}} \sqrt{\teps - \frac 83} \left[1 - \frac{39}{256}\left(\teps - \frac 83\right) + \mathcal O\f{\left(\teps - \frac 83\right)^2}\right], \\
  \abs{\mathcal R} = \frac{3\sqrt{3}}{8\sqrt{2}} \sqrt{\teps - \frac 83} \left[1 - \frac{51}{256}\left(\teps - \frac 83\right) + \mathcal O\f{\left(\teps - \frac 83\right)^2}\right]
\end{gather*}
at $\teps \to 8/3 + 0$;
\begin{gather*}
  \begin{multlined}
    \teps_{\mathrm{SN}}\f{\tomega} = \frac 83 + \frac{4}{\sqrt{3}} \left(\tomega - \frac{1}{\sqrt{3}}\right) - \frac 14 \left(\tomega - \frac{1}{\sqrt{3}}\right)^2 \\
    {}+ \mathcal O\f{\left(\tomega - \frac{1}{\sqrt{3}}\right)^3},
  \end{multlined} \\
  \begin{multlined}
    \tau = \frac{1}{\sqrt{3}} + \frac{1}{4} \left(\tomega - \frac{1}{\sqrt{3}}\right) - \frac{15\sqrt{3}}{128} \left(\tomega - \frac{1}{\sqrt{3}}\right)^2 \\
    {}+ \mathcal O\f{\left(\tomega - \frac{1}{\sqrt{3}}\right)^3},
  \end{multlined} \\
  \begin{multlined}
    \psi = \frac{\pi}{3} + \frac{3}{8} \left(\tomega - \frac{1}{\sqrt{3}}\right) - \frac{51\sqrt{3}}{256} \left(\tomega - \frac{1}{\sqrt{3}}\right)^2 \\
    {}+ \mathcal O\f{\left(\tomega - \frac{1}{\sqrt{3}}\right)^3},
  \end{multlined} \\
  \varrho = \frac{3\sqrt{3}}4 \left(\tomega - \frac{1}{\sqrt{3}}\right) - \frac{27}{128} \left(\tomega - \frac{1}{\sqrt{3}}\right)^2 + \mathcal O\f{\left(\tomega - \frac{1}{\sqrt{3}}\right)^3}, \\
  \begin{multlined}
    \rho = \frac{3^{3/4}}{2\sqrt{2}} \sqrt{\tomega - \frac{1}{\sqrt{3}}} \\
    {}\times \left[1 - \frac{15\sqrt{3}}{64}\left(\tomega - \frac{1}{\sqrt{3}}\right) + \mathcal O\f{\left(\tomega - \frac{1}{\sqrt{3}}\right)^2}\right],
  \end{multlined} \\
  \begin{multlined}
    \abs{\mathcal R} = \frac{3^{5/4}}{4\sqrt{2}} \sqrt{\tomega - \frac{1}{\sqrt{3}}} \\
    {}\times \left[1 - \frac{19\sqrt{3}}{64}\left(\tomega - \frac{1}{\sqrt{3}}\right) + \mathcal O\f{\left(\tomega - \frac{1}{\sqrt{3}}\right)^2}\right]
  \end{multlined}
\end{gather*}
 at $\tomega \to 1/\sqrt{3} + 0$;
\begin{gather*}
  \tomega_{\mathrm{SN}}\f{\teps} = \frac{\teps}{2} - 1 + \frac{1}{\teps} - \frac{2}{\teps^2} + \mathcal O\f{\frac{1}{\teps^3}}, \\
  \tau = 1 - \frac{2}{\teps} + \frac{6}{\teps^2} - \frac{28}{\teps^3} + \mathcal O\f{\frac{1}{\teps^4}}, \\
  \psi = \frac{\pi}2 - \frac{2}{\teps} + \frac{4}{\teps^2} - \frac{52}{3\teps^3} + \mathcal O\f{\frac{1}{\teps^4}}, \\
  \varrho = \frac{\teps}{2} - 1 - \frac{2}{\teps} + \frac{8}{\teps^2} + \mathcal O\f{\frac{1}{\teps^3}}, \\
  \rho = 1 - \frac{2}{\teps} + \frac{2}{\teps^2} - \frac{12}{\teps^3} + \mathcal O\f{\frac{1}{\teps^4}}, \\
  \abs{\mathcal R} = \frac 1{\sqrt{2}} \left(1 - \frac{1}{\teps} - \frac{5}{2\teps^2} + \frac{11}{2\teps^3} + \mathcal O\f{\frac{1}{\teps^4}}\right)
\end{gather*}
at $\teps \to \infty$;
\begin{gather*}
  \teps_{\mathrm{SN}}\f{\tomega} = 2\tomega + 2 - \frac{1}{\tomega} + \frac{2}{\tomega^2} + \mathcal O\f{\frac{1}{\tomega^3}}, \\
  \tau = 1 - \frac{1}{\tomega} + \frac{5}{2\tomega^2} - \frac{8}{\tomega^3} + \mathcal O\f{\frac{1}{\tomega^4}}, \\
  \psi = \frac{\pi}2 - \frac{1}{\tomega} + \frac{2}{\tomega^2} - \frac{17}{3\tomega^3} + \mathcal O\f{\frac{1}{\tomega^4}}, \\
  \varrho = \tomega - \frac{3}{2\tomega} + \frac{4}{\tomega^2} + \mathcal O\f{\frac{1}{\tomega^3}}, \\
  \rho = 1 - \frac{1}{\tomega} + \frac{3}{2\tomega^2} - \frac{4}{\tomega^3} + \mathcal O\f{\frac{1}{\tomega^4}}, \\
  \abs{\mathcal R} = \frac{1}{\sqrt{2}}\left[1 - \frac{1}{2\tomega} - \frac{1}{8\tomega^2} + \frac{19}{16\tomega^3} + \mathcal O\f{\frac{1}{\tomega^4}}\right]
\end{gather*}
at $\tomega \to \infty$.
A fun note is that the coefficient in expansion for $\tomega_{\mathrm{SN}}(\teps)$ at $\tomega \to \infty$ starting with term $1/\varepsilon$ are integer with alternating signs, and their absolute values form OEIS sequence A196148 $(1, 2, 7, 30, \ldots)$, which occurs in combinatorial and statistical problems on random walks over two-dimensional lattices~\cite{kreweras_solution_1981, wood_renyi_2017, oeis_a196148}.

\subsubsection{Partially synchronized states at small and large couplings}
\label{app:symmetric:couplings}

Let us present the asymptotics for parameters of the partially synchronized states at large and small couplings.
At $\teps \to 2(\tomega^2 + 1) + 0$ for the symmetric state, one has
\begin{gather*}
  \begin{multlined}
    \varrho = \frac{\teps - 2\tomega^2 - 2}{1 - 3\tomega^2} - \frac{4\tomega^2 \left(\teps- 2\tomega^2 - 2\right)^2}{\left(1- 3\tomega^2\right)^3} \\
    {}+ \frac{\tomega^2 \left(17 \tomega^2 + 5\right) \left(\teps - 2\tomega^2 - 2\right)^3}{\left(1- 3\tomega^2\right)^5} + \mathcal O\f{\left(\teps - 2\tomega^2 - 2\right)^4},
  \end{multlined} \\
  \begin{multlined}
    \abs{\mathcal R} = \sqrt{\frac{\teps - 2\tomega^2 - 2}{2\left(\tomega^2 + 1\right)\left(1 - 3\tomega^2\right)}} \\
    {}\times \twolinebrackets{[}{]}{1 - \frac{\left(17\tomega^4 + 2\tomega^2 + 1\right) \left(\teps- 2\tomega^2 - 2\right)}{4\left(\tomega^2 + 1\right)\left(1- 3\tomega^2\right)^2}}%
    {{}+ \mathcal O\f{\left(\teps - 2\tomega^2 - 2\right)^2}},
  \end{multlined} \\
  \begin{multlined}
    \tau = \tomega - \frac{\tomega \left(\teps- 2\tomega^2 - 2\right)}{1- 3\tomega^2} \\
    {}+ \frac{\tomega \left(\tomega^2 + 1\right) \left(\teps - 2\tomega^2 - 2\right)^2}{\left(1- 3\tomega^2\right)^3} + \mathcal O\f{\left(\teps - 2\tomega^2 - 2\right)^3},
  \end{multlined} \\
  \begin{multlined}
    \psi = 2\arctan \tomega - \frac{2\tomega \left(\teps- 2\tomega^2 - 2\right)}{\left(\tomega^2 + 1\right)\left(1- 3\tomega^2\right)} \\
    {}+ \frac{2\tomega \left(4\tomega^4 + \tomega^2 + 1\right) \left(\teps - 2\tomega^2 - 2\right)^2}{\left(\tomega^2 + 1\right)^2\left(1- 3\tomega^2\right)^3} \\
    {}+ \mathcal O\f{\left(\teps - 2\tomega^2 - 2\right)^3},
  \end{multlined} \\
  \begin{multlined}
  \end{multlined}
    \rho = \sqrt{\frac{\teps - 2\tomega^2 - 2}{2\left(1 - 3\tomega^2\right)}} \twolinebrackets{[}{]}{1 - \frac{\left(5 \tomega^2 + 1\right) \left(\teps - 2\tomega^2 - 2\right)}{4\left(1- 3\tomega^2\right)^2}}%
    {{}+ \mathcal O\f{\left(\teps - 2\tomega^2 - 2\right)^2}}.
\end{gather*}
At $\teps \to \infty$, one has
\begin{gather*}
  \varrho = \teps - 2 - \frac{\tomega^2}{\teps} + \mathcal O\f{\teps^{-2}}, \\
  \abs{\mathcal R} = 1 - \frac{1}{\teps} - \frac{\tomega^2 + 1}{2\teps^2} + \mathcal O\f{\teps^{-3}}, \\
  \tau = \frac{\tomega}{\teps} + \frac{\tomega}{\teps^2} + \frac{\tomega \left(\tomega^2 + 1\right)}{\teps^3} + \mathcal O\f{\teps^{-4}}, \\
  \psi = \frac{2\tomega}{\teps} + \frac{2\tomega}{\teps^2} + \frac{2\tomega \left(2\tomega^2 + 3\right)}{3\teps^3} + \mathcal O\f{\teps^{-4}}, \\
  \rho = 1 - \frac{1}{\teps} - \frac{1}{2\teps^2} - \frac{2\tomega^2 + 1}{2\teps^3} + \mathcal O\f{\teps^{-4}}.
\end{gather*}

At $\teps \to 4 + 0$ for the asymmetric states, one obtains
\begin{gather*}
  \varrho = \frac{\teps - 4}{4} - \frac{\left(\teps - 4\right)^2}{64\tomega^2} - \frac{\left(2\tomega^2 + 3\right)\left(\teps - 4\right)^3}{512\tomega^4} + \mathcal O\f{\left(\teps - 4\right)^4}, \\
  \psi = \frac{\pi}{2} + \frac{\teps - 4}{8\tomega} + \frac{3\left(\teps - 4\right)^2}{128\tomega^3} + \mathcal O\f{\left(\teps - 4\right)^3}, \\
  \begin{multlined}
    \tOmega_s = \pm \sqrt{\tomega^2 - 1} \twolinebrackets{[}{]}{1 - \frac{3\left(\teps - 4\right)}{8\left(\tomega^2 - 1\right)}}%
    { {}- \frac{\left(4\tomega^2 - 1\right)\left(2\tomega^2 + 1\right)\left(\teps - 4\right)^2}{128 \tomega^2\left(\tomega^2 - 1\right)^2} + \mathcal O\f{\left(\teps - 4\right)^3}},
  \end{multlined} \\
  \begin{multlined}
    \frac{\rho_2}{\rho_1} = \left(\tomega \pm \sqrt{\tomega^2 - 1}\right) \\
    {}\times \left[1 \mp \frac{\left(2\tomega^2 + 1\right) \left(\teps - 4\right)}{8\tomega \sqrt{\tomega^2 - 1}} + \mathcal O\f{\left(\teps - 4\right)^2}\right],
  \end{multlined} \\
  \begin{multlined}
    \abs{\mathcal R} = \frac{\sqrt{\teps - 4}}{4} \twolinebrackets{[}{]}{1 - \frac{\left(4\tomega^2 + 1\right)\left(\teps - 4\right)}{32\tomega^2}}%
    { {}+ \frac{\left(48\tomega^4 - 8\tomega^2 - 25\right)\left(\teps - 4\right)^2}{2048\tomega^4} + \mathcal O\f{\left(\teps - 4\right)^3}},
  \end{multlined} \\
    \rho_2 = \sqrt{\frac{\left(\tomega \pm \sqrt{\tomega^2 - 1}\right) \left(\teps - 4\right)}{8\tomega}} \\
    {}\times \left[1 + \frac{3\sqrt{\tomega^2 - 1}\mp2\tomega \left(2\tomega^2 + 1\right)}{32\tomega^2\sqrt{\tomega^2 - 1}}\left(\teps - 4\right) + \mathcal O\f{\left(\teps - 4\right)^2}\right].
  \begin{multlined}
  \end{multlined}
\end{gather*}

\subsection{Asymmetric case with different subgroup populations}
\label{app:asymmetric1}

In all formulas for $\rho_{1, 2}$, the upper sign in $\pm$, $\mp$ refers to $\rho_1$, and the lower one corresponds to $\rho_2$.
In indices, $(1)$, $(-1)$ are equivalent to $(+)$ and $(-)$, respectively.

\subsubsection{The first Andronov–Hopf bifurcation of the incoherent state}
\label{app:asymmetric1:ah1}

The asymptotics for $\teps_{\mathrm{AH}1}(\tomega, \Dg)$ at $\omega \to 0$ and $\omega \to \infty$ are given, respectively, by
\begin{multline*}
  \teps_{\mathrm{AH}1}\f{\tomega, \Dg} = 2 + 2\left(1 - \Dg^2\right) \tomega^2 - 8 \Dg^2 \left(1 - \Dg^2\right) \tomega^4 \\
  {}- 16 \Dg^2 \left(1 - \Dg^2\right) \left(1 - 3\Dg^2\right) \tomega^6 + \mathcal O\f{\tomega^8}
\end{multline*}
and
\begin{multline*}
  \teps_{\mathrm{AH}1}\f{\tomega, \Dg} = \frac{4}{1 + \abs{\Dg}} - \frac{2 \abs{\Dg} \left(1 - \abs{\Dg}\right)}{\left(1 + \abs{\Dg}\right)^3 \tomega^2} \\
  {}- \frac{\abs{\Dg} \left(1 - \abs{\Dg}\right) \left(3 - 6\abs{\Dg} - \Dg^2\right)}{2\left(1 + \abs{\Dg}\right)^5 \tomega^4} + \mathcal O\f{\frac 1{\tomega^6}}.
\end{multline*}
Note that the substitution $\abs{\Dg} \mapsto -\abs{\Dg}$ transforms the last equality into the asymptotics for the second Andronov–Hopf bifurcation of the incoherent state (where the unstable partially synchronized state is born and the incoherent state remains unstable on both sides of bifurcation).

\subsubsection{Saddle-node and saddle-node infinite-period bifurcations}
\label{app:asymmetric1:sn}

The position of the first saddle-node bifurcation or saddle-node infinite period bifurcation and the respective partially synchronized states satisfy the following asymptotics at $\Dg \to 0$:
\begin{gather*}
  \begin{multlined}
    \teps_{\mathrm{SN}1}\f{\tomega, \Dg} = \teps_{\mathrm{SN}} - 4 \left(1 - \tau_{\mathrm{SN}}^2\right) \Dg^2 \\
    {}+ \frac{8}{\teps_{\mathrm{SN}}} \left(68 - 11\tau_{\mathrm{SN}}^2 + 2\tau_{\mathrm{SN}}^4 - \frac{15}{\tau_{\mathrm{SN}}^2} - \frac{96}{3 - \tau_{\mathrm{SN}}^2} \right) \Dg^4 \\
    {}+ \mathcal O\f{\Dg^6},
  \end{multlined} \\
  \begin{multlined}
    \varrho = \frac{3\tau_{\mathrm{SN}}^2 - 1}{1 - \tau_{\mathrm{SN}}^2} - \left[1 + \frac{2\left(3 - 5\tau_{\mathrm{SN}}^2\right)\left(7 - 5\tau_{\mathrm{SN}}^2\right)}{\left(1 - \tau_{\mathrm{SN}}^4\right)\left(3 - \tau_{\mathrm{SN}}^2\right)}\right] \Dg^2 \\
    {}+ \mathcal O\f{\Dg^4},
  \end{multlined}\\
  \begin{multlined}
    \abs{\mathcal R} = \frac{\sqrt{3\tau_{\mathrm{SN}}^2 - 1}}{1 + \tau_{\mathrm{SN}}^2} + \twolinebrackets{[}{]}{\frac{13}2 - \frac{14}{1 + \tau_{\mathrm{SN}}^2} + \frac{32\tau_{\mathrm{SN}}^2}{\left(1 + \tau_{\mathrm{SN}}^2\right)^3}}%
    {{}- \frac{6}{3 - \tau_{\mathrm{SN}}^2}} \frac{\Dg^2}{\sqrt{3\tau_{\mathrm{SN}}^2 - 1}} + \mathcal O\f{\Dg^4},
  \end{multlined} \\
  \begin{multlined}
    \tOmega_s = \frac{\teps_{\mathrm{SN}}\tau_{\mathrm{SN}}}{1 + \tau_{\mathrm{SN}}^2} \Dg - \frac{4}{\teps_{\mathrm{SN}}\tau_{\mathrm{SN}}} \twolinebrackets{[}{]}{65 + 10\tau_{\mathrm{SN}}^2 + \tau_{\mathrm{SN}}^4}%
    {{}+ \frac{48}{3 - \tau_{\mathrm{SN}}^2} + \frac{256 \left(18 - 5\tau_{\mathrm{SN}}^2\right)}{\left(7 - \tau_{\mathrm{SN}}^2\right)^2}} \Dg^3 + \mathcal O\f{\Dg^5},
  \end{multlined} \\
  \begin{multlined}
    \rho_{1, 2} = \sqrt{\frac{3\tau_{\mathrm{SN}}^2 - 1}{1 + \tau_{\mathrm{SN}}^2}} \twolinebrackets{[}{]}{1 \mp \frac{1 - \tau_{\mathrm{SN}}^2}{1 + \tau_{\mathrm{SN}}^2}\Dg + \frac{1 - \tau_{\mathrm{SN}}^2}2}%
    {{}\times \left(\frac{19 - 3\tau_{\mathrm{SN}}^2}{\left(1 + \tau_{\mathrm{SN}}^2\right)^2} - \frac{3}{3 - \tau_{\mathrm{SN}}^2} -\frac{7}{3\tau_{\mathrm{SN}}^2 - 1}\right)\Dg^2 + \mathcal O\f{\Dg^3}},
  \end{multlined} \\
  \psi = 2\arctan\tau_{\mathrm{SN}} + \frac{4 \left(9 - 11\tau_{\mathrm{SN}}^2\right)}{\teps_{\mathrm{SN}}\tau_{\mathrm{SN}} \left(3 - \tau_{\mathrm{SN}}^2\right)} \Dg^2 + \mathcal O\f{\Dg^4},
\end{gather*}
where the argument $\tomega$ is omitted at $\teps_{\mathrm{SN}}(\tomega)$ and $\tau_{\mathrm{SN}}(\tomega)$ for brevity,
\begin{gather*}
  \tau_{\mathrm{SN}}\f{\tomega} = \sqrt{\frac{\sqrt{\teps_{\mathrm{SN}} \left(\teps_{\mathrm{SN}} + 8\right)} - \teps_{\mathrm{SN}} - 2}2}
\end{gather*}
is the tangent of $\psi/2$ at the saddle-node bifurcation in the symmetric case, and $\teps_{\mathrm{SN}}(\tomega)$ is given by Eq.~\eqref{eq:eps-sn}.
One can also obtain asymptotics at $\tomega \to \infty$,
\begin{gather*}
  \begin{multlined}
    \teps_{\mathrm{SN}1}\f{\tomega, \Dg} = 2\tomega + 2 -\frac{1 + 6\Dg^2 + \Dg^4}{\left(1 - \Dg^2\right)^2 \tomega} \\
    {}+ \frac{2 \left(1 + \Dg^2\right) \left(1 + 7\Dg^2 + 23\Dg^4 + \Dg^6\right)}{\left(1 - \Dg^2\right)^2 \tomega^2} + \mathcal O\f{\frac 1{\tomega^3}},
  \end{multlined} \\
  \begin{multlined}
    \varrho = \left(1 + \Dg^2\right) \tomega + 2\Dg^2 \\
    {}- \frac{3 + 45\Dg^2 + 61\Dg^4 + 3\Dg^6}{2\left(1 - \Dg^2\right)^2 \tomega} + \mathcal O\f{\frac 1{\tomega^2}},
  \end{multlined} \\
  \begin{multlined}
    \abs{\mathcal R} = \sqrt{\frac{1 + \Dg^2}2} - \frac{1 - \Dg^2}{2 \sqrt{2\left(1 + \Dg^2\right)} \,\tomega} \\
    {}- \frac{1 + 84\Dg^2 + 190\Dg^4 + 100\Dg^6 + 9\Dg^8}{8\sqrt{2}\, \left(1 + \Dg^2\right)^{3/2} \left(1 - \Dg^2\right)^2 \tomega^2} + \mathcal O\f{\frac 1{\tomega^3}},
  \end{multlined} \\
  \begin{multlined}
    \tOmega_s = \Dg \tomega + \Dg - \frac{\Dg \left(1 + 6\Dg^2 + \Dg^4\right)}{\left(1 - \Dg^2\right)^2 \tomega} + \mathcal O\f{\frac 1{\tomega^2}},
  \end{multlined} \\
  \begin{multlined}
    \rho_{1, 2} = 1 - \frac{1}{\left(1 \mp \Dg\right)\tomega} + \frac{3 \mp 3\Dg + 4\Dg^2}{2\left(1 \mp \Dg\right)^3\tomega^2} + \mathcal O\f{\frac 1{\tomega^3}},
  \end{multlined} \\
  \begin{multlined}
    \psi = \frac{\pi}2 - \frac{1 + \Dg^2}{\left(1 - \Dg^2\right) \tomega} + \frac{2 \left(1 + 9\Dg^2 + 13\Dg^4 + \Dg^6\right)}{\left(1 - \Dg^2\right)^3 \tomega^2} \\
    {}+ \mathcal O\f{\frac 1{\tomega^3}}.
  \end{multlined}
\end{gather*}

For the second saddle-node bifurcation, one has
\begin{gather*}
  \begin{multlined}
    \teps_{\mathrm{SN}2}\f{\tomega, \Dg} = \left(\tomega^{\frac 23} + 1\right)^2 - \tomega^{\frac 49} \left(\tomega^{\frac 23} + 1\right)^{\frac 43} \\
    {}\times \left(\tomega^{\frac 43} - 4 \tomega^{\frac 23} + 7\right)^{\frac 13} \left(\frac{\abs{\Dg}}2\right)^{\frac 23} + \frac 13 \tomega^{\frac 29} \left(\tomega^{\frac 23} + 1\right)^{\frac 23} \\
    {}\times \left(3\tomega^{\frac{10}3} - 14\tomega^{\frac 83} + 27\tomega^2 + 12\tomega^{\frac 43} - 128\tomega^{\frac 23} + 192\right) \\
    {}\times \left(\tomega^{\frac 43} - 4 \tomega^{\frac 23} + 7\right)^{-\frac 43} \left(\frac{\abs{\Dg}}2\right)^{\frac 43} + \mathcal O\f{\Dg^2},
  \end{multlined} \\
  \begin{multlined}
    \varrho = \tomega^{\frac 23} - 1 + \tomega^{\frac 49} \left(\tomega^{\frac 23} + 1\right)^{\frac 13} \left(\tomega^{\frac 43} - 3 \tomega^{\frac 23} + 4\right) \\
    {}\times \left(\tomega^{\frac 43} - 4 \tomega^{\frac 23} + 7\right)^{-\frac 23}\left(\frac{\abs{\Dg}}2\right)^{\frac 23} + \mathcal O\f{\abs{\Dg}^{\frac 43}},
  \end{multlined} \\
  \begin{multlined}
    \abs{\mathcal R} = \frac{\sqrt{\tomega^{\frac 23} - 1}}{\tomega^{\frac 23} + 1} \\
    {}+ \frac{\tomega^{\frac 49} \left(2 \tomega^2 - 7\tomega^{\frac 43} + 12\tomega^{\frac 23} - 3\right)}{2 \left(\tomega^{\frac 23} + 1\right)^{\frac 53} \left(\tomega^{\frac 43} - 4 \tomega^{\frac 23} + 7\right)^{\frac 23} \sqrt{\tomega^{\frac 23} - 1}} \left(\frac{\abs{\Dg}}2\right)^{\frac 23} \\
    {}+ \mathcal O\f{\abs{\Dg}^{\frac 43}},
  \end{multlined} \\
  \begin{multlined}
    \tOmega_s = \frac{\tomega^{\frac 59}\left(\tomega^{\frac 23} + 1\right)^{\frac 23}}{\left(\tomega^{\frac 43} - 4 \tomega^{\frac 23} + 7\right)^{\frac 13}} \sqrt[3]{\frac{\Dg}2} + \twolinebrackets{[}{]}{5\tomega^{\frac 23} + 4}%
    {{}+ \frac{2\left(17\tomega^{\frac 23} - 52\right)}{\tomega^{\frac 43} - 4 \tomega^{\frac 23} + 7} - \frac{12\left(7\tomega^{\frac 23} - 11\right)}{\left(\tomega^{\frac 43} - 4 \tomega^{\frac 23} + 7\right)^2}} \frac{\tomega^{\frac 13}\Dg}{6} \\
    {}+ \mathcal O\f{\abs{\Dg}^{\frac 53}},
  \end{multlined} \\
  \begin{multlined}
    \rho_{1, 2} = \frac{\left(\tomega^{\frac 23} - 1\right)^{\frac 12}}{\left(\tomega^{\frac 23} + 1\right)^{\frac 12}} \mp \frac{\tomega^{\frac 23} \left(\tomega^{\frac 23} - 1\right)^{\frac 12}}{\left(\tomega^{\frac 23} + 1\right)^{\frac 56} \left(\tomega^{\frac 43} - 4 \tomega^{\frac 23} + 7\right)^{\frac 13}} \sqrt[3]{\frac{\Dg}2} \\
    {}- \frac{\tomega^{\frac 43} \left(3\tomega^{\frac 23} - 7\right)}{2\left(\tomega^{\frac 23} - 1\right)^{\frac 12} \left(\tomega^{\frac 23} + 1\right)^{\frac 76} \left(\tomega^{\frac 43} - 4 \tomega^{\frac 23} + 7\right)^{\frac 23}} \left(\frac{\abs{\Dg}}2\right)^{\frac 23} \\
    {}+ \mathcal O\f{\Dg},
  \end{multlined} \\
  \begin{multlined}
    \psi = 2\arctan \tomega^{\frac 13} \\
    {}- \frac{2\tomega^{\frac 13} \left(\tomega^{\frac 43} - 3\tomega^{\frac 23} +5\right)}{\left(\tomega^{\frac 23} + 1\right)^{\frac 23} \left(\tomega^{\frac 43} - 4 \tomega^{\frac 23} + 7\right)^{\frac 23}} \left(\frac{\abs{\Dg}}2\right)^{\frac 23} + \mathcal O\f{\abs{\Dg}^{\frac 43}}.
  \end{multlined}
\end{gather*}

\subsubsection{The ‘incoherent’ Bogdanov–Takens points and the cusp point}
\label{app:asymmetric1:ibt}

The asymptotics for the positions of two Bogdanov–Takens points with $\varrho = 0$ at $\Dg \to 0$ are as follows:
\begin{gather*}
  \tomega_{\mathrm{SN}1}\f{\Dg} = \frac 1{\sqrt{3}} \left[1 + \frac{28\Dg^2}3 + \frac{1210\Dg^4}3 + \mathcal O\f{\Dg^6}\right], \\
  \begin{multlined}
    \teps_{\mathrm{SN}1}\f{\tomega_{\mathrm{SN}1}\f{\Dg}, \Dg} = \frac 83 + \frac{88 \Dg^2}9 + \frac{13160 \Dg^4}{27} \\
    {}+ \mathcal O\f{\Dg^6},
  \end{multlined} \\
  \begin{multlined}
    \tomega_{\mathrm{SN}2}\f{\Dg} = 1 - \frac 32 \left(\frac{\abs{\Dg}}2\right)^{2/3} - \frac{29}{8} \left(\frac{\abs{\Dg}}2\right)^{4/3} - \frac{55 \Dg^2}{64} \\
    {}+ \mathcal O\f{\abs{\Dg}^{8/3}},
  \end{multlined} \\
  \begin{multlined}
    \teps_{\mathrm{SN}2}\f{\tomega_{\mathrm{SN}2}\f{\Dg}, \Dg} = 4 - 8 \left(\frac{\abs{\Dg}}{2}\right)^{2/3} + \frac{8}{3} \left(\frac{\abs{\Dg}}2\right)^{4/3} \\
    {}- \frac{4 \Dg^2}{3} + \mathcal O\f{\abs{\Dg}^{8/3}}.
  \end{multlined}
\end{gather*}
One can also obtain the asymptotic expressions for the cusp point and the respective parameters of the partially synchronized state at $\abs{\Dg} \to 1$,
\begin{gather*}
  \begin{multlined}
    \teps_{\mathrm{C}}\f{\Dg} = \frac{4}{\left(1 - \abs{\Dg}\right)^2} + \frac{2}{1 - \abs{\Dg}} - \frac{4}{3} - \abs{\Dg} \\
    {}+ \mathcal O\f{\left(1 - \abs{\Dg}\right)^2},
  \end{multlined} \\
  \tomega_{\mathrm{C}}\f{\Dg} = \frac{\teps_{\mathrm{C}}\f{\Dg}}2 - \frac 34 - \frac{1 - \abs{\Dg}}8 + \mathcal O\f{\left(1 - \abs{\Dg}\right)^2}, \\
  \varrho = \abs{\Dg}\teps_{\mathrm{C}}\f{\Dg} + \frac{4 \left(1 - \abs{\Dg}\right)}3 + \mathcal O\f{\left(1 - \abs{\Dg}\right)^2}, \\
  \begin{multlined}
    \abs{\mathcal R} = 1 - \frac{1 - \abs{\Dg}}{2} - \frac{\left(1 - \abs{\Dg}\right)^2}{8} + \frac{5\left(1 - \abs{\Dg}\right)^3}{48} \\
    {}+ \mathcal O\f{\left(1 - \abs{\Dg}\right)^4},
  \end{multlined} \\
  \tOmega_s = \Dg\tomega_{\mathrm{C}}\f{\Dg} + \sgn \Dg \frac{1 + 2\abs{\Dg}}3 + \mathcal O\f{\left(1 - \abs{\Dg}\right)^2}, \\
  \rho_{\frac{3 - \sgn\Dg}2} = 1 - \frac{1 - \abs{\Dg}}2 + \frac{5\left(1 - \abs{\Dg}\right)^2}{24} + \mathcal O\f{\left(1 - \abs{\Dg}\right)^2}, \\
  \begin{multlined}
    \rho_{\frac{3 + \sgn\Dg}2} = 1 - \frac{\left(1 - \abs{\Dg}\right)^2}4 - \frac{23\left(1 - \abs{\Dg}\right)^4}{96} \\
    {}+ \mathcal O\f{\left(1 - \abs{\Dg}\right)^5},
  \end{multlined} \\
  \begin{multlined}
    \psi = \frac{\pi}2 - \frac{\left(1 - \abs{\Dg}\right)^2}{12} - \frac{\left(1 - \abs{\Dg}\right)^3}{36} + \mathcal O\f{\left(1 - \abs{\Dg}\right)^4}.
  \end{multlined} 
\end{gather*}

\subsubsection{The Andronov–Hopf bifurcation of the coherent state}
\label{app:asymmetric1:ah2}

The asymptotics at $\tomega \to \infty$ are as follows:
\begin{gather*}
  \begin{multlined}
    \teps_{\mathrm{AH}2}\f{\tomega, \Dg} = \frac{4}{1 - \abs{\Dg}} + \frac{2\abs{\Dg}\left(1 + 3\abs{\Dg}\right)}{\left(1 - \abs{\Dg}\right)^3 \tomega^2} \\
    + \frac{\abs{\Dg}\left(3 + 17\abs{\Dg} + 77\Dg^2 + 15\abs{\Dg}^3\right)}{2\left(1 - \abs{\Dg}\right)^5 \tomega^4} + \mathcal O\f{\frac 1{\tomega^6}},
  \end{multlined} \\
  \begin{multlined}
    \varrho = \frac{\abs{\Dg}\left(1 + \abs{\Dg}\right)}{1 - \abs{\Dg}} \twolinebrackets{[}{]}{2 + \frac{1 + \abs{\Dg} + 2\Dg^2}{\left(1 - \abs{\Dg}\right)^2 \tomega^2}}%
    {{}+ \frac{3 + 14\abs{\Dg} + 38\Dg^2 + 46\abs{\Dg}^3 + 11\Dg^4}{4\left(1 - \abs{\Dg}\right)^4 \tomega^4} + \mathcal O\f{\frac 1{\tomega^6}}},
  \end{multlined} \\
  \begin{multlined}
    \abs{\mathcal R} = \sqrt{\frac{\abs{\Dg}\left(1 + \abs{\Dg}\right)}2} \twolinebrackets{[}{]}{1 + \frac{1 + \abs{\Dg}}{4\left(1 - \abs{\Dg}\right)\tomega^2}}%
    {{}+ \frac{5 + 23\abs{\Dg} + 55\Dg^2 - 3\abs{\Dg}^3}{32\left(1 - \abs{\Dg}\right)^3\tomega^4} + \mathcal O\f{\frac 1{\tomega^6}}},
  \end{multlined} \\
  \begin{multlined}
    \tOmega_s = \sgn \Dg \tomega - \sgn \Dg \frac{1 + 3\abs{\Dg}}{2\left(1 - \abs{\Dg}\right)\tomega} \\
    {}- \sgn \Dg \frac{\left(1 + \abs{\Dg}\right) \left(1 + 12\abs{\Dg} + 27\Dg^2\right)}{8\left(1 - \abs{\Dg}\right)^3 \tomega^3} + \mathcal O\f{\frac 1{\tomega^5}},
  \end{multlined} \\
  \begin{multlined}
    \rho_{\frac{3 - \sgn\Dg}2} = \frac{\sqrt{\abs{\Dg}\left(1 + \abs{\Dg}\right)}}{\sqrt{2} \left(1 - \abs{\Dg}\right) \tomega} \threelinebrackets{[}{]}{1 + \frac{3 + 14\abs{\Dg} + 7\Dg^2}{8\left(1 - \abs{\Dg}\right)^2\tomega^2}}%
    {{}+ \frac{31 + 324\abs{\Dg} + 1178\Dg^2 + 1220\abs{\Dg}^3 + 63\Dg^4}{128\left(1 - \abs{\Dg}\right)^4\tomega^4}}%
    {{}+ \mathcal O\f{\frac 1{\tomega^6}}},
  \end{multlined} \\
  \begin{multlined}
    \rho_{\frac{3 + \sgn\Dg}2} = \sqrt{\frac{2\abs{\Dg}}{1 + \abs{\Dg}}} \twolinebrackets{[}{]}{1 + \frac{1 + 3\abs{\Dg}}{8\left(1 - \abs{\Dg}\right)\tomega^2}}%
    {{}+ \frac{11 + 47\abs{\Dg} + 273\Dg^2 - 11\abs{\Dg}^3}{128\left(1 - \abs{\Dg}\right)^3\tomega^4} + \mathcal O\f{\frac 1{\tomega^6}}},
  \end{multlined} \\
  \begin{multlined}
    \psi = \frac{\pi}2 + \frac{\Dg^2}{2\left(1 - \abs{\Dg}\right)^2 \tomega^3} + \frac{\abs{\Dg}^3 \left(15 + \abs{\Dg}\right)}{4\left(1 - \Dg^2\right)^4 \tomega^5} + \mathcal O\f{\frac 1{\tomega^7}}.
  \end{multlined}
\end{gather*}

\subsubsection{The ‘coherent’ Bogdanov–Takens point}
\label{app:asymmetric1:cbt}

The asymptotic expressions for the Bogdanov–Takens point with nonzero $\varrho$ at $\Dg \to 0$ are as follows,
\begin{gather*}
  \teps_{\mathrm{BT}}\f{\Dg} = 4 + 36 \left(\frac{\abs{\Dg}}2\right)^{4/3} + 57\Dg^2 + \mathcal O\f{\abs{\Dg}^{8/3}}, \\
  \begin{multlined}
    \tomega_{\mathrm{BT}}\f{\Dg} = 1 + \frac 32 \left(\frac{\abs{\Dg}}2\right)^{2/3} + \frac{99}8 \left(\frac{\abs{\Dg}}2\right)^{4/3} \\
    {}+ \frac{1583\Dg^2}{64} + \mathcal O\f{\abs{\Dg}^{8/3}},
  \end{multlined} \\
  \varrho = 2\left(\frac{\abs{\Dg}}2\right)^{2/3} + 12 \left(\frac{\abs{\Dg}}2\right)^{4/3} + \frac{39\Dg^2}{2} + \mathcal O\f{\abs{\Dg}^{8/3}}, \\
  \abs{\mathcal R} = \frac{\abs{\Dg}^{1/3}}{2^{5/6}} + \frac{3\abs{\Dg}}{2^{3/2}} + \frac{21\abs{\Dg}^{5/3}}{2^{19/6}} + \mathcal O\f{\abs{\Dg}^{7/3}}, \\
  \tOmega_s = \sqrt[3]{\frac{\Dg}2} \left[1 - \frac 12 \left(\frac{\abs{\Dg}}2\right)^{2/3} + \frac{99}8 \left(\frac{\abs{\Dg}}2\right)^{4/3} + \mathcal O\f{\Dg^2}\right], \\
  \begin{multlined}
    \rho_{1, 2} = \left(\frac{\abs{\Dg}}2\right)^{1/3} \mp \frac{\sgn\Dg}2 \left(\frac{\abs{\Dg}}2\right)^{2/3} + \frac{19\abs{\Dg}}{16} \\
    {}+ \mathcal O\f{\abs{\Dg}^{4/3}}, 
  \end{multlined} \\
  \psi = \frac{\pi}2 - \left(\frac{\abs{\Dg}}2\right)^{2/3} + \frac 72 \left(\frac{\abs{\Dg}}2\right)^{4/3} + \mathcal O\f{\Dg^2}.
\end{gather*}

Let us also give the asymptotics at $\abs{\Dg} \to 1$,
\begin{gather*}
  \begin{multlined}
    \teps_{\mathrm{BT}}\f{\Dg} = \frac{16\left(\sqrt{2} - 1\right)}{\left(1 - \abs{\Dg}\right)^2} - \frac{4\left(5\sqrt{2} - 7\right)}{1 - \abs{\Dg}} - 9 + \frac{25\sqrt{2}}{4} \\
    {}- \frac{4 + 17\sqrt{2}}{16} \left(1 - \abs{\Dg}\right) + \mathcal O\f{\left(1 - \abs{\Dg}\right)^2},
  \end{multlined} \\
  \tomega_{\mathrm{BT}}\f{\Dg} = \frac{\teps_{\mathrm{BT}}\f{\Dg}}2 - 1 + \frac{1 - \abs{\Dg}}{4\sqrt{2}} + \mathcal O\f{\left(1 - \abs{\Dg}\right)^2}, \\
  \begin{multlined}
    \varrho = \abs{\Dg}\teps_{\mathrm{BT}}\f{\Dg} - 4 + 2\left(2\sqrt{2} - 1\right)\abs{\Dg} + \mathcal O\f{\left(1 - \abs{\Dg}\right)^2},
  \end{multlined} \\
  \begin{multlined}
    \abs{\mathcal R} = 1 - \frac{1 - \abs{\Dg}}{2} - \frac{1 + \sqrt{2}}{16}\left(1 - \abs{\Dg}\right)^2 \\
    {}+ \frac{5 + \sqrt{2}}{64} \left(1 - \abs{\Dg}\right)^3 + \mathcal O\f{\left(1 - \abs{\Dg}\right)^4},
  \end{multlined} \\
  \tOmega_s = \Dg {\tomega_{\mathrm{C}}\f{\Dg} + 2\Dg + \mathcal O\f{\left(1 - \abs{\Dg}\right)^2}}, \\
  \begin{multlined}
    \rho_{\frac{3 - \sgn\Dg}2} = 1 - \frac{1 + \sqrt{2}}4 \left(1 - \abs{\Dg}\right) \\
    {}+ \frac{7 + 2\sqrt{2}}{32} \left(1 - \abs{\Dg}\right)^2 + \mathcal O\f{\left(1 - \abs{\Dg}\right)^3},
  \end{multlined} \\
  \begin{multlined}
    \rho_{\frac{3 + \sgn\Dg}2} = 1 - \frac{1 + \sqrt{2}}{16} \left(1 - \abs{\Dg}\right)^2 \\
    {}- \frac{1 + 3\sqrt{2}}{64} \left(1 - \abs{\Dg}\right)^3 + \mathcal O\f{\left(1 - \abs{\Dg}\right)^4},
  \end{multlined} \\
  \begin{multlined}
    \psi = \frac{\pi}2 + \frac{1 - \abs{\Dg}}{4} - \frac{1 + 2\sqrt{2}}{16}\left(1 - \abs{\Dg}\right)^2 \\
    {}+ \mathcal O\f{\left(1 - \abs{\Dg}\right)^3}.
  \end{multlined} 
\end{gather*}

Interestingly, the angle $\psi$ in synchronized state at the Bogdanov–Takens point approaches $\pi/2$ both at $\Dg \to 0$ and $\abs{\Dg} \to 1$.
Actually, this angle is always close to $\pi/2$, $-0.042 < \psi - \pi/2 < 0.079$  for any asymmetry.

\subsubsection{Partially synchronized states at small and large couplings}
\label{app:asymmetric1:couplings}

Let us present the asymptotics for parameters of the partially synchronized states at large and small couplings.
At $\teps \to \teps_{\mathrm{AH}1}(\tomega, \Dg) + 0$, one has
\begin{gather*}
  \varrho = a_1 \left(\teps - \teps_{\mathrm{AH}1}\right) + a_2 \left(\teps - \teps_{\mathrm{AH}1}\right)^2 + \mathcal O\f{\left(\teps - \teps_{\mathrm{AH}1}\right)^3}, \\
  \begin{multlined}
    \abs{\mathcal R} = b_1 \left(\teps - \teps_{\mathrm{AH}1}\right)^{1/2} + b_2 \left(\teps - \teps_{\mathrm{AH}1}\right)^{3/2} \\
    {}+ \mathcal O\f{\left(\teps - \teps_{\mathrm{AH}1}\right)^{5/2}},
  \end{multlined} \\
  \tOmega_s = c_1 + c_2 \left(\teps - \teps_{\mathrm{AH}1}\right) + \mathcal O\f{\left(\teps - \teps_{\mathrm{AH}1}\right)^2}, \\
  \psi = d_1 \left(\teps - \teps_{\mathrm{AH}1}\right) + d_2 \left(\teps - \teps_{\mathrm{AH}1}\right)^2 + \mathcal O\f{\left(\teps - \teps_{\mathrm{AH}1}\right)^3}, \\
  \begin{multlined}
    \frac{\rho_{1, 2}}{\abs{\mathcal R}} = e_1^{(\pm)} + e_2^{(\pm)} \left(\teps - \teps_{\mathrm{AH}1}\right) + \mathcal O\f{\left(\teps - \teps_{\mathrm{AH}1}\right)^{2}},
  \end{multlined}
\end{gather*}
where the arguments $\tomega$ and $\Dg$ are omitted at $\teps_{\mathrm{AH}1}(\tomega, \Dg)$ and other functions for brevity,
\begin{gather*}
  \begin{multlined}
    a_1 = \frac{2  \left[16\tomega^2 + \left(4 - \teps_{\mathrm{AH}1}\right)^2\right]}{\teps_{\mathrm{AH}1}^2 \zeta_{\mathrm{AH}1}} \\
    {}\times \left[16\left(\tomega^2 + 1\right) - \teps_{\mathrm{AH}1} \left(\teps_{\mathrm{AH}1} + 4\right)\right],
  \end{multlined} \\
  \begin{multlined}
    a_2 = \frac{8\left(\teps_{\mathrm{AH}1} - 2\right) \left[16\tomega^2 + \left(4 - \teps_{\mathrm{AH}1}\right)^2\right]}{\teps_{\mathrm{AH}1}^3 \zeta_{\mathrm{AH}1}^3} \\
    {}\times\fivelinebrackets{[}{]}{4096\left(\tomega^2 + 1\right)^3 \left(10\tomega^2 + 11\right) - 1024 \left(\tomega^2 + 1\right)^2 \left(12\tomega^4\right.}%
    {\left.{} + 65\tomega^2 + 57\right)\teps_{\mathrm{AH}1} + 512\left(\tomega^2 + 1\right)\left(31\tomega^4 + 86\tomega^2 + 57\right)}%
    {{}\times\teps_{\mathrm{AH}1}^2 + 128\left(\tomega^2 + 1\right)\left(4\tomega^4 + 57\tomega^2 + 62\right)\teps_{\mathrm{AH}1}^3 + 16}%
    {{}\times\left(14\tomega^4 + 93\tomega^2 + 83\right)\teps_{\mathrm{AH}1}^4 + 4\left(4\tomega^4 + 3\tomega^2 - 19\right)\teps_{\mathrm{AH}1}^5}%
    {{}- 6\left(\tomega^2 + 2\right)\teps_{\mathrm{AH}1}^6 + \teps_{\mathrm{AH}1}^7},
  \end{multlined} \\
  b_1 = \sqrt{\frac{a_1}{\teps_{\mathrm{AH}1}}}, \quad b_2 = \frac{\teps_{\mathrm{AH}1}a_2 - a_1}{2\sqrt{\teps_{\mathrm{AH}1} a_1}}, \\
  c_1 = \tOmega_{\mathrm{AH}1}, \\
  c_2 = -\frac{2\tOmega_{\mathrm{AH}1} \left(\teps_{\mathrm{AH}1} - 2\right) \left[48\left(\tomega^2 + 1\right) - \teps_{\mathrm{AH}1} \left(\teps_{\mathrm{AH}1} + 8\right)\right]}{\teps_{\mathrm{AH}1}\zeta_{\mathrm{AH}1}}, \\
  \begin{multlined}
    d_1 = \psi_{\mathrm{AH}1} = \arctan \frac{4\tomega}{4 - \teps_{\mathrm{AH}1}} \\
    {}= \arctan\f{\tomega - \tOmega_{\mathrm{AH}1}} + \arctan\f{\tomega + \tOmega_{\mathrm{AH}1}},
  \end{multlined} \\
  d_2 = -\frac{8\tomega \left[4\left(\tomega^2 + 1\right) \left(\teps_{\mathrm{AH}1} - 2\right) - \teps_{\mathrm{AH}1}^2\right]}{\teps_{\mathrm{AH}1}\zeta_{\mathrm{AH}1}}, \\
  e_1^{(\pm)} = \frac{\teps_{\mathrm{AH}1}}{2 \sqrt{\left(\tomega \pm \tOmega_{\mathrm{AH}1}\right)^2 + 1}}, \\
  \begin{multlined}
    e_2 = \frac{8 \left(\tomega \pm \tOmega_{\mathrm{AH}1}\right)}{\zeta_{\mathrm{AH}1}\sqrt{\left(\tomega \pm \tOmega_{\mathrm{AH}1}\right)^2 + 1}} \twolinebrackets{[}{]}{\left(\tomega \mp \tOmega_{\mathrm{AH}1}\right)}%
    {{}\times \left(\tomega^4 - \tOmega_{\mathrm{AH}1}^4 - 1\right) \pm \tOmega_{\mathrm{AH}1} \left(3\tomega^2 + \tOmega_{\mathrm{AH}1}^2 + 3\right)},
  \end{multlined} \\
  \zeta_{\mathrm{AH}1}\f{\tomega, \Dg} = 8\left(\tomega^2 + 1\right)\left(16\tomega^2 + 24 - 18\teps_{\mathrm{AH}1} + \teps_{\mathrm{AH}1}^2\right) \\
  {}+ \teps_{\mathrm{AH}1}^2 \left(20 - \teps_{\mathrm{AH}1}\right).
\end{gather*}
Since the above expressions are for coefficients quite cumbersome, let us also give their asymptotics at $\tomega \to 0$,
\begin{gather*}
  a_1 = 1 + 3 \left(1 - \Dg^2\right) \tomega^2 + \mathcal O\f{\tomega^4}, \\
  a_2 = -4 \left(1 - \Dg^2\right) \tomega^2 + \mathcal O\f{\tomega^4}, \\
  b_1 = \frac{1}{\sqrt{2}} + \frac{1 - \Dg^2}{\sqrt{2}}\tomega^2 + \mathcal O\f{\tomega^4}, \\
  b_2 = -\frac{1}{4\sqrt{2}} - \sqrt{2} \left(1 - \Dg^2\right) \tomega^2 + \mathcal O\f{\tomega^4}, \\
  c_1 = \Dg\tomega + 2\Dg\left(1 - \Dg^2\right) \tomega^3 + \mathcal O\f{\tomega^5}, \\
  c_2 = -7\Dg \left(1 - \Dg^2\right) \tomega^3 + \mathcal O\f{\tomega^5}, \\
  d_1 = 2\tomega - \frac{2\left(1 + 3\Dg^2\right) \tomega^3}3 + \mathcal O\f{\tomega^5}, \\
  d_2 = -2\tomega - 4 \left(1 - 5\Dg^2\right) \tomega^3 + \mathcal O\f{\tomega^5}, \\
  e_1^{(\pm)} = 1 + \frac{\left(1 \pm \Dg\right)\left(1 \mp 3\Dg\right)\tomega^2}2 + \mathcal O\f{\tomega^4}, \\
  e_2^{(\pm)} = -\left(1 \pm \Dg\right)\left(1 \mp 4\Dg\right)\tomega^2 + \mathcal O\f{\tomega^4},
\end{gather*}
and $\tomega \to \infty$,
\begin{gather*}
  a_1 = \frac{\left(1 + \abs{\Dg}\right)^2}4 + \frac{3\abs{\Dg} \left(1 - \abs{\Dg}\right)}{8 \tomega^2} + \mathcal O\f{\frac 1{\tomega^4}}, \\
  a_2 = - \frac{\left(1 -\Dg^2\right)\left(1 - 5\abs{\Dg}\right)}{64 \tomega^2} + \mathcal O\f{\frac 1{\tomega^4}}, \\
  b_1 = \frac{\left(1 + \abs{\Dg}\right)^{3/2}}4 + \frac{\abs{\Dg} \left(1 - \abs{\Dg}\right)}{4 \sqrt{1 + \abs{\Dg}}\, \tomega^2} + \mathcal O\f{\frac 1{\tomega^4}}, \\
  b_2 = -\frac{\left(1 + \abs{\Dg}\right)^{5/2}}{32} - \frac{\left(1 - \abs{\Dg}\right) \left(1 + \abs{\Dg}\right)^{3/2}}{128 \tomega^2} + \mathcal O\f{\frac 1{\tomega^4}}, \\
  c_1 = \sgn \Dg \tomega - \sgn \Dg \frac{1 - \abs{\Dg}}{2\left(1 + \abs{\Dg}\right) \tomega} + \mathcal O\f{\frac 1{\tomega^3}}, \\
  c_2 = -\sgn \Dg \frac{3\left(1 - \abs{\Dg}\right)}{8 \tomega} + \mathcal O\f{\frac 1{\tomega^3}}, \\
  d_1 = \frac{\pi}2 - \frac{\abs{\Dg}}{\left(1 + \abs{\Dg}\right) \tomega} + \mathcal O\f{\frac 1{\tomega^3}}, \\
  d_2 = \frac{1 - \abs{\Dg}}{8\tomega} + \mathcal O\f{\frac 1{\tomega^3}}, \\
  e_1^{(\sgn\Dg)} = \frac{1}{\left(1 + \abs{\Dg}\right) \tomega} + \frac{1 - 6\abs{\Dg} + \Dg^2}{8\left(1 + \abs{\Dg}\right)^3 \tomega^3} + \mathcal O\f{\frac 1{\tomega^5}}, \\
  e_1^{(-\sgn\Dg)} = \frac{2}{1 + \abs{\Dg}} - \frac{\left(1 - \abs{\Dg}\right)\left(1 + 3 \abs{\Dg}\right)}{4 \left(1 + \abs{\Dg}\right)^3 \tomega^2} + \mathcal O\f{\frac 1{\tomega^4}}, \\
  e_2^{(\sgn\Dg)} = \frac{1}{4\tomega} + \frac{9 - 7\abs{\Dg}}{32\left(1 + \abs{\Dg}\right) \tomega^3} + \mathcal O\f{\frac 1{\tomega^5}}, \\
  e_2^{(-\sgn\Dg)} = -\frac{\abs{\Dg}\left(1 - \abs{\Dg}\right)}{4 \left(1 + \abs{\Dg}\right)^2 \tomega^2} + \mathcal O\f{\frac 1{\tomega^4}}.
\end{gather*}

At $\teps \to \infty$, one obtains
\begin{gather*}
  \begin{multlined}
    \varrho = \teps - 2 - \frac{\left(1 - \Dg^2\right)\tomega^2}{\teps} - \frac{2\left(1 - \Dg^2\right)\tomega^2}{\teps^2} \\
    {}- \frac{\left(1 - \Dg^2\right)\tomega^2\left(\tomega^2 + 3\right)}{\teps^3} \\
    {}- \frac{2\left(1 - \Dg^2\right)\tomega^2\left[\left(3 - \Dg^2\right)\tomega^2 + 2\right]}{\teps^4} + \mathcal O\f{\frac 1{\teps^5}},
  \end{multlined} \\
  \begin{multlined}
    \abs{\mathcal R} = 1 - \frac 1{\teps} - \frac{\left(1 - \Dg^2\right)\tomega^2 + 1}{2\teps^2} - \frac{3\left(1 - \Dg^2\right)\tomega^2 + 1}{2\teps^3} \\
    {}+ \mathcal O\f{\frac 1{\teps^4}},
  \end{multlined} \\
  \begin{multlined}
    \tOmega_s = \Dg\tomega + \frac{\Dg\left(1 - \Dg^2\right)\tomega^3}{\teps^3} + \frac{4\Dg\left(1 - \Dg^2\right)\tomega^3}{\teps^4} \\
    {}+ \mathcal O\f{\frac 1{\teps^5}},
  \end{multlined} \\
  \rho_{1, 2} = 1 - \frac 1{\teps} - \frac{1}{2\teps^2} - \frac{2\left(1 \pm \Dg\right)\tomega^2 + 1}{2\teps^3} + \mathcal O\f{\frac 1{\teps^4}}, \\
  \begin{multlined}
    \psi = \frac{2\tomega}{\teps} + \frac{2\tomega}{\teps^2} + \frac{2\tomega \left(2\tomega^2 + 3\right)}{3\teps^3} + \frac{2\tomega \left[\left(3 - \Dg^2\right)\tomega^2 + 1\right]}{\teps^4} \\
    {}+ \mathcal O\f{\frac 1{\teps^5}}.
  \end{multlined}
\end{gather*}

\subsubsection{Partially synchronized states at small and large frequency differences}
\label{app:asymmetric1:omega}

At $\tomega \to 0$, the bimodal distribution turns into unimodal one with one stable partially synchronized state existing at $\teps > 2$.
The respective asymptotics are as follows:
\begin{gather*}
  \begin{multlined}
    \varrho = \teps - 2 - \frac{\left(1 - \Dg^2\right)\teps\tomega^2}{\left(\teps - 1\right)^2} \\
    {}- \frac{\left(1 - \Dg^2\right)\teps\left[\teps^2 - 2\Dg^2\teps - 1 - 3\Dg^2\right]\tomega^4}{\left(\teps - 1\right)^6} + \mathcal O\f{\tomega^6},
  \end{multlined} \\
  \abs{\mathcal R} = \sqrt{1 - \frac{2}{\teps}} \left[1 - \frac{\left(1 - \Dg^2\right)\teps\tomega^2}{2\left(\teps - 1\right)^2\left(\teps - 2\right)} + \mathcal O\f{\tomega^4}\right], \\
  \tOmega_s = \Dg\tomega + \frac{\Dg \left(1 - \Dg^2\right) \teps\tomega^3}{\left(\teps - 1\right)^4} + \mathcal O\f{\tomega^5}, \\
  \rho_{1, 2} = \sqrt{1 - \frac{2}{\teps}} \left[1 - \frac{\left(1 \pm \Dg\right)\left(\teps - 1 \mp \Dg\right)\tomega^2}{\left(\teps - 1\right)^3\left(\teps - 2\right)} + \mathcal O\f{\tomega^4}\right], \\
  \psi = \frac{2\tomega}{\teps - 1} + \frac{2\left[2\teps^2 - \left(1 + 3\Dg^2\right)\left(\teps + 1\right)\right]\tomega^3}{3\left(\teps - 1\right)^5} + \mathcal O\f{\tomega^5}.
\end{gather*}

At $\tomega \to \infty$, there are two finite stationary partially synchronized solutions $\varrho_{\pm}$ existing at $\teps > 4/(1 \pm \abs{\Dg})$.
The asymptotics of the solution $\varrho_{+}$ are given by following expressions:
\begin{gather*}
  \begin{multlined}
    \varrho = \frac{\left(1 + \abs{\Dg}\right)\left[\left(1 + \abs{\Dg}\right)\teps - 4\right]}4 \\
    {}- \frac{\left(1 + \abs{\Dg}\right)\teps\left\{\left(1 + \abs{\Dg}\right)\left[\left(1 - \abs{\Dg}\right)\teps - 4\right]^2 - 32\abs{\Dg}\right\}}{256\tomega^2} \\
    {}+ \mathcal O\f{\frac 1{\tomega^4}},
  \end{multlined} \\
  \begin{multlined}
    \abs{\mathcal R} = \frac{1 + \abs{\Dg}}2 \sqrt{1 - \frac{4}{\left(1 + \abs{\Dg}\right)\teps}} \\
    {}\times\twolinebrackets{[}{]}{1 - \frac{\teps\left\{\left(1 + \abs{\Dg}\right)\left[\left(1 - \abs{\Dg}\right)\teps - 4\right]^2 - 32\abs{\Dg}\right\}}{128\left[\left(1 + \abs{\Dg}\right)\teps - 4\right]\tomega^2}}%
    {{}+ \mathcal O\f{\frac 1{\tomega^4}}},
  \end{multlined} \\
  \begin{multlined}
    \tOmega_s = \sgn \Dg \tomega - \sgn\Dg \frac{\left(1 - \abs{\Dg}\right)\teps\left[\left(1 + \abs{\Dg}\right)\teps - 2\right]}{16\tomega} \\
    {}+ \mathcal O\f{\frac 1{\tomega^3}},
  \end{multlined} \\
  \begin{multlined}
    \frac{\rho_{\frac{3 - \sgn\Dg}2}}{\abs{\mathcal{R}}} = \frac{\teps}{4\tomega} + \frac{\teps\left[\left(1 + \abs{\Dg}\right)\left(3 - \abs{\Dg}\right)\teps^2 - 8\left(\teps + 1\right)\right]}{256\tomega^3} \\
    {}+ \mathcal O\f{\frac 1{\tomega^5}},
  \end{multlined} \\
  \begin{multlined}
    \frac{\rho_{\frac{3 + \sgn\Dg}2}}{\abs{\mathcal{R}}} = \frac{2}{1 + \abs{\Dg}} + \frac{\left(1 - \abs{\Dg}\right)\teps\left[\left(1 - \abs{\Dg}\right)\teps - 8\right]}{64 \left(1 + \abs{\Dg}\right)\tomega^2} \\
    {}+ \mathcal O\f{\frac 1{\tomega^4}},
  \end{multlined} \\
  \psi = \frac{\pi}2 + \frac{\left(1 - \abs{\Dg}\right) \teps - 4}{8\tomega} + \mathcal O\f{\frac 1{\tomega^3}}.
\end{gather*}
The respective expressions for the solution $\varrho_{-}$ can be obtained from the above asymptotics for $\varrho = \varrho_{+}$ by substituting $\abs{\Dg} \mapsto -\abs{\Dg}$, $\sgn{\Dg} \mapsto -\sgn{\Dg}$.

\bibliographystyle{apsrev4-2}
\bibliography{bimodal.bib}

\end{document}